\RequirePackage[l2tabu,orthodox]{nag}
\documentclass
[11pt,letterpaper]
{article}

\usepackage[dvipsnames]{xcolor}
\PassOptionsToPackage{pagebackref}{hyperref}

\usepackage[notes=true,later=false,camera=false]{dtrt}

\usepackage[utf8]{inputenc}
\usepackage{etex}
\usepackage{ stmaryrd }
\usepackage{xspace,enumerate}
\usepackage[T1]{fontenc}
\usepackage[full]{textcomp}
\usepackage[american]{babel}
\usepackage{mathtools}

\usepackage{amsthm}
\usepackage{empheq}

   \usepackage{hyperref}
   \hypersetup{
colorlinks=true,
urlcolor=Cerulean,
linkcolor=RoyalBlue,
citecolor=PineGreen,
linktocpage=true,
}

\usepackage{paralist}
\usepackage{turnstile}
\usepackage{mdframed}
\usepackage{tikz}
\usepackage{caption}
\DeclareCaptionType{Algorithm}
\usepackage{newfloat}

\usepackage[capitalise,nameinlink]{cleveref}
\usepackage{thm-restate}

\crefname{lemma}{Lemma}{Lemmas}
\crefname{fact}{Fact}{Facts}
\crefname{theorem}{Theorem}{Theorems}
\crefname{mtheorem}{Theorem}{Theorems}
\crefname{corollary}{Corollary}{Corollaries}
\crefname{claim}{Claim}{Claims}
\crefname{example}{Example}{Examples}
\crefname{algorithm}{Algorithm}{Algorithms}
\crefname{problem}{Problem}{Problems}
\crefname{definition}{Definition}{Definitions}
\crefname{question}{Question}{Questions}
\usepackage{newfloat}
\newtheorem{theorem}{Theorem}[section]
\newtheorem{mtheorem}{Theorem}
\newtheorem*{theorem*}{Theorem}

\newtheorem{proposition}[theorem]{Proposition}
\newtheorem*{proposition*}{Proposition}

\newtheorem*{lemma*}{Lemma}

\newtheorem*{conjecture*}{Conjecture}
\newtheorem{fact}[theorem]{Fact}
\newtheorem*{fact*}{Fact}

\newtheorem*{hypothesis*}{Hypothesis}

\theoremstyle{definition}
\newtheorem{definition}[theorem]{Definition}
\newtheorem*{definition*}{Definition}

\theoremstyle{remark}
\newtheorem{claim}[theorem]{Claim}
\newtheorem*{claim*}{Claim}

\newtheorem*{remark*}{Remark}
\newtheorem{observation}[theorem]{Observation}
\newtheorem*{observation*}{Observation}

\usepackage[
letterpaper,
top=1.2in,
bottom=1.2in,
left=1in,
right=1in]{geometry}
\usepackage{newpxtext} 
\usepackage{textcomp} 
\usepackage[varg,bigdelims]{newpxmath}
\usepackage[scr=rsfso]{mathalfa}
\usepackage{bm} 
\let\mathbb\varmathbb
\usepackage{microtype}
\usepackage{multirow}

\allowdisplaybreaks
\newcommand{\FormatAuthor}[3]{
\begin{tabular}{c}
#1 \\ {\small\texttt{#2}} \\ {\small #3}
\end{tabular}
}

\newcommand{\R}{{\mathbb R}}
\newcommand{\N}{{\mathbb N}}
\newcommand{\norm}[1]{\lVert #1 \rVert}

\newcommand{\abs}[1]{\lvert #1 \rvert}

\newcommand{\eps}{\varepsilon}
\renewcommand{\epsilon}{\eps}
\newcommand{\defeq}{\coloneq}
\newcommand{\F}{{\mathbb F}}

\DeclareMathOperator{\E}{\mathbb{E}}

\newcommand{\ip}[1]{\langle #1 \rangle}

\newcommand{\Bits}{\{0,1\}}

\newcommand{\cH}{\mathcal H}

\newcommand{\ceil}[1]{\lceil #1 \rceil}

\newcommand{\mper}{\,.}
\newcommand{\mcom}{\,,}

\newcommand{\Norm}[1]{\left\lVert#1\right\rVert}

\newcommand{\Id}{\mathrm{I}}

\newcommand{\polylog}{\operatorname{polylog}}

\newcommand{\cT}{\mathcal T}

\newcommand{\cV}{\mathcal{V}}

\newcommand{\Dec}{\operatorname{Dec}}

\newcommand{\Code}{C}
\newcommand{\vspan}{\operatorname{span}}

\newcommand{\cW}{\mathcal{W}}

\newcommand{\bfv}{\mathbf{v}}
\newcommand{\bbfv}{\bar{\mathbf{v}}}
\newcommand{\davg}{d_{\mathrm{avg}}}
\newcommand{\cdeg}{c}
\newcommand{\dout}{d_{\mathrm{out}}}
\newcommand{\din}{d_{\mathrm{in}}}
\newcommand{\cK}{\mathcal{K}}

\newcommand{\Ener}{\mathcal{E}}
\newcommand{\colourset}{\mathscr{C}}

\newcommand{\drbw}{d_{\mathrm{rbw}}}

\begin{document}

\title{Reed--Muller Codes are Optimal Binary Linear $3$-Query LCCs}
\author{
\begin{tabular}[h!]{ccc}
     \FormatAuthor{Oliver Janzer}{oliver.janzer@epfl.ch}{EPFL} &
     \FormatAuthor{Peter Manohar\thanks{Supported by NSF Grant DMS-2424441.}}{pmanohar@ias.edu}{The Institute for Advanced Study} 
\end{tabular}
} %
\date{\today}


\maketitle

\begin{abstract}
We prove that any linear $3$-query locally correctable code $\Code \colon \F_2^k \to \F_2^n$ with constant correction radius must have length $n \geq 2^{\Omega(\sqrt{k})}$. This improves on the prior best lower bound of $2^{\Omega(\sqrt{k/\log k})}$, due to~\cite{AlrabiahG24}, and shows that the folklore construction of linear $3$-LCCs from Reed--Muller codes, which achieve length $n = 2^{O(\sqrt{k})}$, is optimal up to constant factors in the exponent.

Our proof builds on the approach of~\cite{KothariM23}, further refined in~\cite{Yankovitz24,KothariM24}, that uses ``chain derived equations'' in combination with Kikuchi graphs. The crux of our argument is to show that a certain edge-coloured Kikuchi graph has a unique colour cycle: a cycle in which some colour is used exactly once. We do this by controlling the distribution of colours at each vertex in the graph using a random walk mixing argument and a novel hypergraph decomposition based on edge expansion of the hypergraph's ``flattening'' into a graph.
\end{abstract}


\thispagestyle{empty}
\clearpage
 \microtypesetup{protrusion=false}
  \tableofcontents{}
  \microtypesetup{protrusion=true}

\thispagestyle{empty}
\clearpage

\pagestyle{plain}
\setcounter{page}{1}
\section{Introduction}
\label{sec:intro}
Locally correctable codes (LCCs) are error-correcting codes that are additionally equipped with a local self-corrector: an algorithm that, when given oracle access to a corrupted codeword, is capable of recovering any symbol of the underlying uncorrupted codeword with good probability while only reading a small number of randomly chosen symbols from the corrupted codeword. Such codes are formally defined as follows.

\begin{definition}[Locally correctable codes]
\label{def:lcc}
A code $\Code \colon \Bits^k \to \Bits^n$ is a $(q, \delta, \eps)$-LCC if there is a randomized decoding algorithm $\Dec(\cdot)$ with the following properties. The algorithm $\Dec(\cdot)$ is given oracle access to a string $y \in \Bits^n$, takes an index $u \in [n]$ as input, and outputs an element of $\Bits$ with the following guarantees:
\begin{enumerate}[(1)]
\item for any $y$ and $u$, the algorithm $\Dec^{y}(u)$ reads at most $q$ indices of $y$,
\item for all $b \in \Bits^k$, $y \in \Bits^n$ with $\Delta(y, \Code(b)) \leq \delta n$, and $u \in [n]$, $\Pr[\Dec^{y}(u) = \Code(b)_u] \geq 1 - \eps$. Here, $\Delta(x,y) \defeq \abs{\{i : x_i \ne y_i\}}$ denotes the Hamming distance between $x$ and $y$.
\end{enumerate}
In this definition, $\delta$ and $\eps$ are typically taken to be small constants. We write $q$-LCC to refer to a $(q, \delta, \eps)$-LCC where $\delta, \eps$ are constant.
\end{definition}

Locally correctable codes are an important class of error-correcting codes that have numerous intriguing applications throughout theoretical computer science. They were first introduced implicitly in works on program checking~\cite{BlumK95}, and were used in constructions of probabilistically checkable proofs (PCPs) and the proof of the PCP theorem~\cite{AroraLMSS98,AroraS98}. Other applications (see the surveys of~\cite{Trevisan04,Yekhanin12,Dvir12} for more details) include sublinear algorithms and property testing~\cite{RubinfeldS96,BlumLR93}, \textsc{IP}=\textsc{PSPACE}~\cite{LundFKN90,Shamir90}, worst-case to average-case reductions~\cite{BabaiFNW93}, constructions of explicit rigid matrices~\cite{Dvir10}, private information retrieval protocols~\cite{IshaiK99,BarkolIW10}, connections to incidence geometry~\cite{Dvir12}, additive combinatorics~\cite{BhowmickDL13}, and the theory of block designs~\cite{BarkolIW10}. 

Because of their importance, constructing better LCCs or understanding their inherent limitations has remained a significant goal in coding theory over the past three decades. The central question of study is to understand the optimal dependence of $n$, the length of the code, as a function both of $k$, the length of the underlying message, and $q$, the number of queries permitted by the decoder; $\delta, \eps \in (0,1)$ are implicitly taken to be small constants. The folklore LCCs based on Reed--Muller codes, or low-degree multivariate polynomials, provide a useful baseline construction. In the regime where $q$ is a constant, degree-$(q-1)$ Reed--Muller codes give a $q$-LCC $\Code \colon \Bits^k \to \Bits^n$ with $n = 2^{O(k^{1/(q-1)})}$. Moreover, this code is \emph{linear}, meaning that the encoding map $\Code$ is an $\F_2$-linear map.
For super-constant $q$, we know that this baseline construction can be improved, as multiplicity codes~\cite{KoppartySY11,KoppartyMRS16,GopiKORS17} give constructions of LCCs that have a much smaller blocklength $n$ than Reed--Muller codes. However, for constant $q$, Reed--Muller codes remain the best-known LCCs, and finding better constructions remains an open question. In fact, the lack of progress on constructing better constant query LCCs has led to the conjecture that Reed--Muller codes may in fact be optimal constant query LCCs.

For the case of $q = 2$, this conjecture is known to be true. The Hadamard code, or equivalently Reed--Muller codes of degree $1$, gives a $2$-query LCC of length $n = 2^k$, and classic works on lower bounds for locally decodable codes\footnote{Local decodability is a weaker notion than local correctability, and only requires that the decoder be capable of recovering the message bits, rather than all codeword bits. For linear codes, it is easy to see that any LCC is an LDC by choosing a systematic encoding map. For nonlinear codes, a similar statement is also true, see~\cite[Appendix A]{BhattacharyyaGT17}.}  show that $n \geq 2^{\Omega(k)}$, initially for linear codes~\cite{GoldreichKST02}, and later for nonlinear codes~\cite{KerenidisW04} as well. For the case of $q = 3$, however, the situation was much less understood until very recently. Reed--Muller codes of degree $2$ give a $3$-LCC with $n = 2^{O(\sqrt{k})}$, while the classic lower bound of~\cite{KerenidisW04} shows that $n \geq \tilde{\Omega}(k^2)$. Up to $\polylog(n)$ factor improvements, this remained the best lower bound until~\cite{AlrabiahGKM23} improved the lower bound to $n \geq \tilde{\Omega}(k^3)$. Then, the work of~\cite{KothariM23} proved that $n \geq 2^{\Omega(k^{1/8})}$ for any linear $3$-LCC, and this lower bound led to a flurry of subsequent improvements~\cite{Yankovitz24,AlrabiahG24,KothariM24}. The current best bound for \emph{linear} $3$-LCCs is $n \geq 2^{\Omega(\sqrt{k/\log k})}$ due to~\cite{AlrabiahG24}, which is tantalizingly close to the $n = 2^{O(\sqrt{k})}$ achieved by Reed--Muller codes, and~\cite{KothariM24} shows a tight bound of $n\geq 2^{\Omega(\sqrt{k})}$ assuming that the linear $3$-LCC is a ``block design'', which is an additional structural property satisfied by Reed--Muller codes. For nonlinear codes,~\cite{KothariM24} shows a weaker bound of $n \geq 2^{\Omega(k^{1/5})}$.

As the main result of this paper, we prove a linear $3$-LCC lower bound of $n \geq 2^{\Omega(\sqrt{k})}$, showing that the folklore construction of Reed--Muller codes is optimal for linear $3$-LCCs.
\begin{mtheorem}
\label{mthm:main}
Let $\Code \colon \F_2^k \to \F_2^n$ be a linear $(3, \delta, \eps)$-LCC for any $0 \leq \eps < 1/2$. Then, $n \geq 2^{\Omega(\delta^7 \sqrt{k})}$. 
\end{mtheorem}
We note that we did not attempt to optimize the dependence on $\delta$ in \cref{mthm:main}.

\cref{mthm:main} improves on the prior best lower bound of~\cite{AlrabiahG24}, which shows a bound of $n \geq 2^{\Omega(\sqrt{k/\log k})}$. However, as we shall discuss in \cref{sec:proofoverview}, the proof of \cref{mthm:main} follows a different approach compared to~\cite{AlrabiahG24}. In~\cite{AlrabiahG24}, they reduce the task of proving a linear $3$-LCC lower bound to the rainbow cycle problem in extremal combinatorics and then apply known bounds~\cite{AlonBSZZ25}. In comparison, the proof of \cref{mthm:main} follows the ``Kikuchi graph'' approach of~\cite{KothariM23}, further refined in~\cite{Yankovitz24,KothariM24}, and directly constructs a certain type of coloured cycle, which we term ``unique colour cycle'', in the Kikuchi graph. The proof of \cref{mthm:main} introduces several key new ideas, which we will discuss next in \cref{sec:proofoverview}.

\section{Proof overview}
\label{sec:proofoverview}

In this section, we give an overview of the key ideas used to prove \cref{mthm:main}, and compare these ideas to the proofs in prior work~\cite{KothariM23,Yankovitz24,AlrabiahG24,KothariM24}. The proof in this paper, like the proofs in all these previous papers, relies on the following standard combinatorial definition of linear $3$-LCCs, which is equivalent to \cref{def:lcc} up to constant factors~\cite{Dvir16}. For the purposes of this proof overview, we will mostly omit the dependence on $\delta$ in \cref{mthm:main}.
\begin{fact}[Linear $3$-LCC, combinatorial definition; Definition 2.3 and Fact 2.4 in~\cite{KothariM23}]
\label{fact:lcccomb}
Let $\Code \colon \F_2^k \to \F_2^n$ be a $(3,\delta,\eps)$-locally correctable code where $\Code$ is an $\F_2$-linear map and $\eps < 1/2$. For each $u \in [n]$, let $\bfv_u$ denote the $u$-th row vector of the map $\Code$, and note that $\dim(\vspan\{\bfv_1, \dots, \bfv_n\}) = k$ since $\Code$ is injective. For $u \in [2n] \setminus [n]$, let $\bfv_u \defeq \mathbf{0}$. Then, for each $u \in [2n]$, there exists a $3$-uniform hypergraph matching $\cH_u \subseteq {[2n] \setminus \{u\} \choose 3}$ with vertex set $[2n] \setminus \{u\}$ such that (1) $\abs{\cH_u} \geq \delta n/3$, i.e., the matching has at least $\delta n/3$ hyperedges, and (2) for every $u \in [2n]$ and every $\{u_1, u_2, u_3\} \in \cH_u$, it holds that $\bfv_u = \bfv_{u_1} + \bfv_{u_2} + \bfv_{u_3}$.
\end{fact}

\cref{fact:lcccomb} implies that in order to prove \cref{mthm:main}, i.e., to show that every linear $(3, \delta, \eps)$-locally correctable code of length $n$ has $k \leq O_{\delta}(\log^2 n)$, it suffices to prove the following statement: for every set of $n$ vectors $\{\bfv_u\}_{u \in [n]}$ in a vector space over $\F_2$ and every collection of hypergraph matchings $\{\cH_u\}_{u \in [n]}$, each of size at least $\delta n/3$, where $\bfv_u = \bfv_{u_1} + \bfv_{u_2} + \bfv_{u_3}$ for all $u \in [n]$ and $\{u_1, u_2, u_3\} \in \cH_u$, it holds that $\dim(\vspan\{\bfv_1, \dots, \bfv_n\}) \leq O_{\delta}(\log^2 n)$.

\parhead{The challenges in improving the analysis of~\cite{AlrabiahG24}.} The current best bound on $\dim(\vspan\{\bfv_1, \dots, \bfv_n\})$ is $O(\log^2 n \log \log n)$, shown by~\cite{AlrabiahG24}. One natural route to proving \cref{mthm:main} is thus to optimize the analysis of~\cite{AlrabiahG24} and remove the $O(\log \log n)$ factor. However, as we will now discuss, this appears to be quite challenging, as it requires an analogous improvement on the rainbow cycle problem in extremal combinatorics.

The rainbow cycle problem asks the following question: suppose we are given a graph $G$ with average degree $\davg$ that is \emph{properly edge-coloured}, namely, each edge is assigned a colour and the colours on the edges adjacent to any single vertex are all distinct. How large does $\davg$ need to be to guarantee that $G$ contains a \emph{rainbow cycle}, a cycle that uses each colour at most once? The rainbow cycle problem has been studied extensively in extremal combinatorics~\cite{KMSV04,DLS13,Jan23,Tom24,JS24,KLLT24,AlonBSZZ25}, and the current best bound is $\drbw \leq O(\log n \log \log n)$~\cite{AlonBSZZ25}, where $\drbw$ denotes the smallest bound on $\davg$ that guarantees the existence of a rainbow cycle. This is tight up to the $O(\log \log n)$ factor, as the Boolean hypercube (where each coordinate direction denotes a distinct colour) is a properly edge-coloured graph of degree $\log_2 n$ with no rainbow cycle.

Using a very clever reduction, the work of~\cite{AlrabiahG24} shows that $\dim(\vspan\{\bfv_1, \dots, \bfv_n\}) \leq \drbw \cdot O_{\delta}(\log n)$, and thus by~\cite{AlonBSZZ25} this implies that $k \leq O_{\delta}(\log^2 n \log \log n)$. To improve their analysis, one thus needs to remove the $O(\log \log n)$ factor from the rainbow cycle bound, which is a challenging task. Another natural approach would be to adapt the analysis of~\cite{AlrabiahG24} to rely on a weaker combinatorial statement than rainbow cycle bounds, ideally one for which an $O(\log n)$ bound would be easier to show. However, the use of rainbow cycle bounds appears intrinsic to the approach of~\cite{AlrabiahG24}, so it is not clear how to do this either.

\parhead{Our approach.} Instead, to prove \cref{mthm:main} we will follow the more ``direct'' approach used in the work~\cite{KothariM23} and its refinements~\cite{Yankovitz24,KothariM24}. As we shall see, this will only require us to show the existence of an ``odd coloured'' cycle in a specific (not properly) edge-coloured Kikuchi graph obtained from the hypergraph matchings in \cref{fact:lcccomb}; an odd coloured cycle is a cycle that uses \emph{some} colour an odd number of times. We will thus analyze this particular Kikuchi graph directly using its specific properties. Before we turn to the key ideas in our proof, we first summarize the ideas of the works~\cite{KothariM23,Yankovitz24,KothariM24} in \cref{sec:priorapproach} below.

\subsection{The prior approach of~\cite{KothariM23,Yankovitz24,KothariM24}}
\label{sec:priorapproach}
The previous proofs of~\cite{KothariM23,Yankovitz24,KothariM24} follow the same blueprint, introduced in~\cite{KothariM23}. Firstly, rather than analyzing the system of constraints in \cref{fact:lcccomb}, they instead analyze a system of \emph{derived} constraints called ``chain derivations'' formed by applying a specific set of transformations to the original equations. Secondly, they form a Kikuchi graph using the derived constraints, and use this to construct a $2$-LDC of length $N = n^{O(\polylog(n))}$. Thus, by applying known $2$-LDC lower bounds, this implies that $k \leq O(\log N) = O(\polylog(n))$, which finishes the proof. Below, we explain each of these key steps in more detail.

\parhead{Chain derived constraints and the bipartite Kikuchi graph.} The key idea of~\cite{KothariM23} is to analyze not the initial system of constraints in \cref{fact:lcccomb}, but a derived system of constraints formed as follows. For each constraint of the form $\bfv_u = \bfv_{u_1} + \bfv_{u_2} + \bfv_{u_3}$, we can pick another constraint of the form $\bfv_{u_3} = \bfv_{u_4} + \bfv_{u_5} + \bfv_{u_6}$ using the hypergraph matching $\cH_{u_3}$. By combining these two constraints, we get the derived constraint $\bfv_u = \bfv_{u_1} + \bfv_{u_2} + \bfv_{u_4} + \bfv_{u_5} + \bfv_{u_6}$. The definition of $r$-chains given below captures the set of constraints formed by repeating this process $r$ times.

\begin{definition}[Chain derived constraints]
Let $r \geq 1$ be a positive integer. An $r$-chain with \emph{head $u$} is a sequence $C = (u_0, a_1, b_1, u_1, a_2, b_2, u_2, \dots, u_{r-1}, a_r, b_r, u_r)$ where (1) $u_i, a_i, b_i \in [n]$ for all choices of $i$, (2) $u_0 = u$, and (3) for all $i = 1, \dots, r$, we have $\{a_i, b_i, u_i\} \in \cH_{u_{i-1}}$. We let $\cH_u^{(r)}$ denote the set of $r$-chains with head $u$. 

Given a chain $C$ with head $u$, we associate it with the vector $\bfv_C \defeq \bfv_{u_r} + \sum_{i = 1}^r \bfv_{a_i} + \sum_{i = 1}^r \bfv_{b_i}$. Note that this does not include the vector $\bfv_u$ for the head of the chain.
\end{definition}

It is straightforward to observe that if all the constraints defined in \cref{fact:lcccomb} are satisfied, then it holds that $\bfv_C = \bfv_u$ for any $r$-chain $C$ with head $u$.

The next step is to define the following Kikuchi graph. 
\begin{definition}[Bipartite chain Kikuchi graph]
\label{def:bipartitechainkikuchi}
Let $\ell$ be a parameter to be determined later. Let $V_L = {[n] \choose \ell} \times [n]$ and let $V_R = {[n] \choose \ell}$. For each element $u \in [n]$, the bipartite Kikuchi graph $\cK_{r, \ell}(u, \cH)$ is defined as follows. The left vertex set is $V_L$ and the right vertex set is $V_R$. For each $r$-chain $C$ with head $u$, and each $(S,u') \in V_L$ and $T \in V_R$, there is an edge $((S,u'), T)$ ``labeled'' by the $r$-chain $C$ if $S = \{a_1, \dots, a_r\} \cup R$ and $T = \{b_1, \dots, b_r\} \cup R$ for some $R\subseteq [n]$ of size $\abs{R} = \ell - r$ disjoint from both $\{a_1,\dots,a_r\}$ and $\{b_1,\dots,b_r\}$, and $u' = u_r$ in the chain $C$.
\end{definition}

Note that in the above definition, $\{a_1, \dots, a_r\}$ must be a set of size $r$ in order for $S$ to have size $\ell$. This implicitly requires the chains to have \emph{distinct} $a_1, \dots, a_r$ (and similarly for $b_1, \dots, b_r$). This is a minor technicality of \cref{def:bipartitechainkikuchi} that does not materially affect the proof in a major way. We note that the Kikuchi graph that we work with when proving \cref{mthm:main} (\cref{def:kikuchi}) does not have this issue.

The Kikuchi graph is motivated by the following observation. For each set $T \subseteq [n]$, we can associate it with the vector $\bfv_T \defeq \sum_{u \in T} \bfv_u$, and similarly for each left vertex $(S,u')$, we associate it with the vector $\bfv_{(S,u')} \defeq \bfv_{S} + \bfv_{u'}$. If we do this, then we observe that for any edge $((S,u'), T)$ in the graph $\cK_{r, \ell}(u, \cH)$, it holds that $\bfv_{(S,u')} + \bfv_T = \bfv_C = \bfv_u$. The Kikuchi graph thus gives a method for converting the larger arity constraints on the vertex set $[n]$ into arity $2$ constraints on the larger vertex sets $V_L$ and $V_R$.

\parhead{Extracting a $2$-LDC from the Kikuchi graph.} To finish the proof,~\cite{KothariM23,Yankovitz24,KothariM24} then show that the $\F_2$-linear encoding map $\Code'$ given by $x \in \F_2^k \mapsto \{\ip{x, \bfv_S + \bfv_{u'}}\}_{(S,u') \in V_L} \cup \{\ip{x, \bfv_T}\}_{T \in V_R}$ gives a $2$-LDC. This follows by showing that each $\cK_{r, \ell}(u, \cH)$ has a matching of size $\Omega(N)$, where $N = \abs{V_L}$. (Technically, this is not possible since the left and right vertex sets are imbalanced as $\abs{V_L} = \abs{V_R} \cdot n$. However, one can easily overcome this issue by duplicating each right vertex $n$ times and splitting its edges roughly equally across all copies. We ignore this technicality for the proof overview.) It is easy to show that if $\cK_{r, \ell}(u, \cH)$ is $d$-regular, then it has a matching of size $e(\cK_{r, \ell}(u, \cH))/2d$, constructed via a simple greedy algorithm, and a similar statement holds if $\cK_{r, \ell}(u, \cH)$ is approximately regular. Hence, the core technical component of all the works~\cite{KothariM23,Yankovitz24,KothariM24} is to show that $\cK_{r,\ell}(u, \cH)$ contains a subgraph $\cK'$ that contains nearly all of the edges of $\cK_{r,\ell}(u, \cH)$ and is approximately regular in the sense that $d_{\max}(\cK') \leq O(\davg(\cK'))$. This step, sometimes called the ``row pruning'' step because it ``prunes'' the rows of the adjacency matrix of the Kikuchi graph, is typically the core technical piece in works using Kikuchi matrices~\cite{GuruswamiKM22,HsiehKM23,AlrabiahGKM23,BasuHKL25,JanzerM25}.

This core technical component is where the proofs in~\cite{KothariM23,Yankovitz24,KothariM24} diverge. The proof in~\cite{KothariM24} shows that almost all vertices in $\cK_{r, \ell}(u, \cH)$ have degree $(1 + o(1))\davg$ for some $r \leq O(\log n)$ and $\ell=2r$, under the assumption that the $3$-LCC is a \emph{design LCC}. A design $3$-LCC is defined by a $4$-uniform hypergraph $\cH \subseteq {[n] \choose 4}$ where each pair $\{w,w'\} \subseteq [n]$ appears in exactly one hyperedge in $\cH$, and the $3$-uniform hypergraph matchings $\cH_u$ are formed by taking all hyperedges in $\cH$ that contain $u$, and then removing $u$. Note that this means that each $\cH_u$ is a \emph{perfect} hypergraph matching on $[n] \setminus \{u\}$. The calculation in~\cite{KothariM24} showing this $(1+o(1))\davg$ bound is very precise, relying on sharp bounds on binomial coefficients, and crucially uses that the hypergraph matchings $\cH_u$ are perfect matchings and each pair $\{w,w'\}$ appears in exactly one hyperedge. Applying known $2$-LDC lower bounds~\cite{GoldreichKST02} shows that $k \leq O(\log N)$, where $N = \abs{V_L} + \abs{V_R} = O(n^{\ell + 1})$. In this case, we have $\ell = O(r) = O(\log n)$, so this implies that $k \leq O(\log^2 n)$.

However, a $3$-LCC need not be a design LCC in general, as (1) the hypergraph matchings $\cH_u$ need not be perfect, and (2) there may be ``heavy pairs'' of elements $\{w,w'\}$ that appear in many hyperedges across all the $\cH_u$'s. As explained in~\cite{KothariM24}, the analysis of~\cite{Yankovitz24} loses an additional $\log n$ factor to handle each of these two irregularities, resulting in a final bound of $k \leq O(\log^4 n)$. The imperfect matchings can cause large fluctuations in the degrees in the Kikuchi graph when $\ell \approx r$ because, e.g., one can form a $3$-LCC with matchings of size $\delta n$ by repeating a design $3$-LCC $1/\delta$ times on disjoint blocks of coordinates. The standard deviation of the size of the intersection of $S$ with each block then makes the degree of a random left vertex $(S,u')$ in the Kikuchi graph vary by roughly a factor $(1+\Theta_{\delta}(\frac{1}{\sqrt{\ell}}))^r$. These fluctuations therefore become bounded by a constant only once one takes $\ell \approx r^2$, rather than the ideal value of $\ell \approx r$, which loses a $\log n$ factor. 

The second $\log n$ factor loss comes from the issue of ``heavy pairs'' $\{w,w'\}$ that appear in many hyperedges across all the $\cH_u$'s. This issue is handled by a complicated ``chain decomposition'' step that forms the other core technical component of~\cite{KothariM23,Yankovitz24}, and is tailored to make the row pruning work.

\subsection{A proof sketch of \cref{mthm:main}}
\label{sec:proofsketch}
We now explain the key ideas behind the proof of \cref{mthm:main}. From the approach outlined in \cref{sec:priorapproach}, we will make use of the first two ideas, the chain derivations and (a slightly modified version of) Kikuchi graphs. However, we will crucially replace the third component, the reduction to a $2$-LDC, with a different combinatorial statement, which we will then prove using several new ideas. The full proof of \cref{mthm:main} is fairly short, and so we will only sketch the main ideas here.

\parhead{Finding odd-coloured cycles in the chain Kikuchi graph.} Let $k = \dim(\vspan\{\bfv_1, \dots, \bfv_n\})$ be the quantity that we wish to bound. Without loss of generality, we may assume that $\bfv_1, \dots, \bfv_k$ are linearly independent. We define a bipartite edge-coloured graph $\cK_{r, \ell}([k], \cH)$ with colour set $[k]$ using the chain Kikuchi graphs from \cref{def:bipartitechainkikuchi}, by simply taking the set of edges in $\cK_{r, \ell}([k], \cH)$ with colour $i$ to be the edges in $\cK_{r, \ell}(i, \cH)$. For simplicity of notation, we will now let $\cK$ be $\cK_{r, \ell}([k], \cH)$.
The following observation, which is somewhat implicit in prior work~\cite{BhattacharyyaCG20,AlrabiahGKM23,KothariM23,Yankovitz24,AlrabiahG24,KothariM24}, is that to bound $k$, it suffices to show the existence of an \emph{odd-coloured} cycle in $\cK$ if the number of colours $k$ is too large.
\begin{observation}
Let $\cK$ be the edge-coloured graph defined above. Suppose that all of the linear equations defined by the matchings $\cH_u$ are satisfied by the vectors $\bfv_1, \dots, \bfv_n$. Then, $\cK$ does not contain an \emph{odd-coloured} cycle, a cycle where some colour is used an odd number of times. 
\end{observation}
The reason this observation is true is straightforward. Suppose $\cK$ did contain an odd-coloured cycle. Then, if we sum the vectors associated to the vertices across all edges in an odd-coloured cycle, we end up with $\mathbf{0}$, since all vertices in the cycle appear in an even number of edges. However, for each edge $((S,u), T)$ with colour $i$, we have that $\bfv_{(S,u)} + \bfv_T = \bfv_i$, using the linear equations defined by the hypergraph matchings. Hence, if the colour $i$ appears an odd number of times in the cycle, we end up with $\bfv_i$ on the right-hand side after summing over all the edges. But, $\bfv_i\not \in \vspan\{\bfv_j: j\in [k]\setminus \{i\}\}$, so the right-hand side cannot be $\mathbf{0}$, and thus no odd-coloured cycle can exist in~$\cK$.

It thus remains to show that if $k$ is too large, then $\cK$ must contain an odd-coloured cycle. In fact, we will show the existence of a \emph{unique colour} cycle, a cycle where some colour $i$ is used exactly once. Note that, while any rainbow cycle is clearly both a unique colour and odd-coloured cycle, we cannot make black-box use of known rainbow cycle bounds here, as such results require that the host graph is \emph{properly} edge-coloured. Given that the average degree of each colour in $\cK$ is large, this is clearly false.

\parhead{Unique colour cycles exist in graphs with similar degree in each colour.} For a vertex $v \in V_L \cup V_R$, let $d_i(v)$ denote the degree of $v$ with colour $i$, $d(v) = \sum_{i = 1}^k d_i(v)$ denote the degree of $v$, and let $\cdeg(v)$ denote the \emph{colour degree} at $v$, which is $\abs{\{i : d_i(v) \geq 1\}}$. By a simple lemma (\cref{fact:colourcyclebound}), for any edge-coloured graph $G$, if $\cdeg(v) \geq k'$ for all vertices $v$ and $G$ contains no unique colour cycle, then $2^{k'} \leq \abs{V}$. This means that if for some $\ell = O(\log n)$ (so that $\abs{V_L} + \abs{V_R} = n^{O(\log n)}$) we can pass to a subgraph in $\cK$ such that all vertices in this subgraph have colour degree $\Omega(k)$, then we will obtain the desired $k \leq O(\log^2 n)$ bound. 

It thus remains to understand the colour degree of a typical vertex in $\cK$. In fact, one way to argue that the colour degree of a typical vertex is large is to simply show that $d_i(v)$ is approximately the same for almost all $i$ and $v$. Indeed, this is precisely what the technical proofs in~\cite{KothariM23,Yankovitz24,KothariM24} show, as they argue that $\cK_i$, the subgraph consisting only of edges of colour $i$, is approximately regular for all $i$ (and it is straightforward to see that the average degree of each $\cK_i$ is also the same). These technical ``approximate regularity'' lemmas can be easily rephrased to show that a typical vertex has $\cdeg(v) \geq \eta k$ for either $\eta = 1/\polylog(n)$ (\cite{KothariM23,Yankovitz24}) or $\eta = \Omega(1)$ (\cite{KothariM24}).

However, crucially for us, it is possible to argue that $\cdeg(v)$ is typically $\Omega(k)$ \emph{without} showing that $d_i(v) \approx d_j(v')$  for typical colours $i$, $j$, and vertices $v,v'$. In fact, the two $\log n$ factor losses in previous proofs, coming from (1) imperfect hypergraph matchings, and (2) the existence of heavy pairs, arise precisely because each of these irregularities causes the approximate regularity condition to cease to be true. For example, for any pair $\{w,w'\}$ of vertices in $[n]$, if there are many hyperedges containing both $w$ and $w'$, then we may expect the total degree of the left vertex $(S,w')$ to be unusually large whenever $w\in S$. This is because there are many valid choices for the vertices $(u_{r-1},b_r)$ in a chain $C = (u_0, a_1, b_1, u_1, a_2, b_2, u_2, \dots, u_{r-1}, a_r, b_r, u_r)$ with $a_r=w$ and $u_r=w'$.

The central point of our argument is that we will control $\cdeg(v)$ by arguing that a typical vertex has $d_i(v) \approx d_j(v)$, i.e., its degree \emph{within} each colour is approximately the same, even if $d_i(v)$ is not close to $d_i(v')$ for a different vertex $v'$. We do this by analyzing the following expected ``local variance'' quantity, defined as
\begin{equation}
\label{eq:localvar}
\cV \defeq \E_{v \sim V_L \sqcup V_R} \left[\E_{i \sim [k]} \left[\left(d_i(v) - \frac{d(v)}{k}\right)^2\right]\right] \mper
\end{equation}
Notice that if $d_i(v) = d_j(v)$ for all colours $i$ and $j$, then $v$ does not contribute to $\cV$. More generally, if $v$ is a vertex with degree $d(v)$ and has $\cdeg(v) \leq \eta k$, then we have
\begin{equation*}
\E_{i \sim [k]}\left[ \left(d_i(v) - \frac{d(v)}{k}\right)^2 \right] \geq \left(\frac{d(v)}{k}\right)^2 \cdot \left(\frac{1}{\eta} - 1\right) \mcom
\end{equation*}
as the variance is minimized by setting $d_i(v) = d(v)/\eta k$ for $\eta k$ choices of $i$, and $0$ otherwise. This means that if $\cdeg(v)$ is small, then $v$ has a large contribution (scaling with $d(v)^2$) to \cref{eq:localvar}. In other words, if $\cV$ is small, then intuitively most vertices must have large $\cdeg(v)$. We make this intuition precise in the proof of \cref{lem:vartocycle}, where we show that if $\cV \leq (\abs{E}/k\abs{V})^2$, then we can find a subgraph where every vertex $v$ in the subgraph has $\cdeg(v) \geq \Omega(k)$.

\parhead{Bounding the local variance via mixing.} To bound $\cV$ in \cref{eq:localvar}, we now need to argue that for a typical vertex $v$, $d_i(v)$ is similar to $d_j(v)$. For reasons that will be explained later, we define our Kikuchi graph $\cK$ (\cref{def:kikuchi}) slightly differently than in previous work (\cref{def:bipartitechainkikuchi}). First of all, the vertex sets are $V_L=[n]^{r+1}$ and $V_R=[n]^{r}$, that is, instead of sets of size $\ell \geq r$, we will consider ordered tuples of length exactly $r$. Our chains will be defined as $C = (u_0, a_1, b_1, u_1, a_2, b_2, u_2, \dots, u_{r-1}, a_r, b_r, u_r)$, where for each $j$, we have either $\{a_{j},b_{j},u_{j}\}\in \cH_{u_{j-1}}$, or $(u_{j-1},a_{j},b_{j},u_{j})=(x,y,y,x)$ for some $x,y\in [n]$. We say that the tuples of the form $(x,y,y,x)$ are ``fake'' tuples, and the ones corresponding to hyperedges are ``real''.
We now take an edge of colour $i\in [k]$ in our Kikuchi graph between $(a_1, \dots, a_r,u_r)\in V_L$ and $(b_1, \dots, b_r)\in V_R$ for each valid chain of the form $(u_0, a_1, b_1, u_1, a_2, b_2, u_2, \dots, u_{r-1}, a_r, b_r, u_r)$ with $u_0=i$. Crucially, we still have $\bfv_{u_{j-1}}+\bfv_{a_{j}}+\bfv_{b_{j}}+\bfv_{u_{j}}=\mathbf{0}$ for every $j$, so our new Kikuchi graph still has no unique colour cycle.

By rearranging terms in \cref{eq:localvar}, we can show that for this new Kikuchi graph, bounding $\cV$ is equivalent to bounding $\E_{\vec{a}}[\|\mathbf{p}_i(\vec{a}) - \mathbf{p}_{i'}(\vec{a})\|_2^2]$, where $\vec{a}=(a_1,\dots,a_r)$ is uniformly random and $\mathbf{p}_i(\vec{a})$ is the probability distribution of the ``tail vertex'' $u_r$ in a random $r$-chain $C = (u_0, a_1, b_1, u_1, a_2, b_2, u_2, \dots, u_{r-1}, a_r, b_r, u_r)$ with $u_0=i$ and given $a_1,\dots,a_r$.

We can view each $r$-chain as being constructed one link at a time, where the sequence of steps $u_0 \to u_1 \to \dots \to u_r$ is simply a random walk where we transition from $u_{j-1}$ to $u_{j}$ as follows: either we pick $u_{j}$ with $\{a_{j},b_{j},u_{j}\}\in \cH_{u_{j-1}}$ for some $b_{j}\in [n]$ (if such $u_{j}$ exists), or we take $u_{j}=u_{j-1}$. By adding a sufficient number of ``fake'' tuples, we can define the chains such that the probability of staying at $u_{j-1}$ will always be at least $1/2$.
Hence, the distributions $\mathbf{p}_i(\vec{a})$ and $\mathbf{p}_{i'}(\vec{a})$ are simply the distribution of the final vertex in this random walk after taking $r$ steps, either starting from the vertex $i$ or the vertex $i'$. We will be able to show that these distributions are close, as long as the directed graph $D_{\cH}$, containing an edge $(u,u')$ for each $\{a,b,u'\}\in \cH_u$, is a good edge expander (see \cref{lem:mixing}).

The case where $D_{\cH}$ is not an edge expander is more complicated. In essence, we will decompose $D_{\cH}$ into $O(1/\delta^2)$ subsets of vertices, where the induced graph on a subset $W$ is an edge expander. This decomposition argument is subtle, and constructing the $j$-th piece $W_j$ of the decomposition requires ``quotienting out'' by the vectors $\bfv_{u}$ for $u \in W_1 \cup \dots \cup W_{j-1}$ and ignoring vertices $u \in [n]$ if their vectors $\bfv_u$ become $\mathbf{0}$ in the quotient space.

In order to apply the mixing argument, we require that the walk $u_0 \to u_1 \to \dots \to u_r$ corresponding to a chain stays in the piece $W$ in the decomposition, even though the other elements of the chain, namely the $a_j$'s and $b_j$'s, can be arbitrary in $[n]$. Each ``real'' step in the chain can be viewed as $\{a_j, b_j, u_j\} \in \cH_{u_{j-1}}$ where $(u_{j-1}, u_j)$ is the edge in $D_{\cH}$ that is within the expanding piece~$W$.

This allows us to bound the dimension of the quotient of each piece $$\vspan\{\bfv_u: u\in W_1 \cup \dots \cup W_j\} / \vspan\{\bfv_u: u\in W_1 \cup \dots \cup W_{j-1}\}$$ by $O(\log^2 n)$, and conclude using that there are only $O(1/\delta^2)$ pieces.

We give the proof of \cref{mthm:main} in \cref{sec:main,sec:expansion}. In \cref{sec:main}, we begin the proof and execute the decomposition into $O(1/\delta^2)$ expanding pieces. Then, in \cref{sec:expansion} we handle the main technical component of the proof, which is the case where $D_{\cH}$ is an edge expander.


\section{Preliminaries}

For a graph $G = (V,E)$ and $v \in V$, we let $d^G(v)$ denote the degree of $v$ in $G$. When $G$ is clear from context, we will use $d(v)$ instead of $d^G(v)$ for simplicity. Graphs in this paper may have self-loops and parallel edges. For sets $S, T \subseteq V$, we let $e^G(S,T)$ be the number of edges of the form $\{v,v'\}$ where $v \in S$ and $v' \in T$. When $G$ is clear from context, we will omit the subscript and instead write $e(S,T)$.

\begin{definition}[Edge-coloured graphs and coloured cycles]
\label{def:colourgraph}
An edge-coloured graph $G$ with colour palette $[k]$ is a collection $(V,E_1, \dots, E_k)$ where $V$ is the vertex set and each $E_i \subseteq {V \choose 2}$ is an edge multiset. The multiset $E_i$ denotes the edges with colour $i$. We let $d_i(v)$ denote the degree of $v$ with colour $i$, i.e., $d_i(v) \defeq d^{G_i}(v)$ where $G_i = (V,E_i)$, and we let $d(v) = \sum_{i = 1}^k d_i(v)$ denote the total degree of $v$. We let $\cdeg(v) \defeq \abs{\{i \in [k] : d_i(v) \geq 1\}}$ denote the \emph{colour degree} of the vertex $v$.

An odd-coloured cycle in $G$ is a cycle where there exists $i \in [k]$ such that the number of edges appearing in the cycle from $E_i$ is odd. A unique colour cycle in $G$ is a cycle where there exists $i \in [k]$ such that exactly one edge from $E_i$ appears in the cycle. A rainbow cycle in $G$ is a cycle where all colours are distinct. Cycles of length two are allowed.
\end{definition}

The following fact shows that every edge-coloured graph with no unique colour cycle and large minimum colour degree must have a large number of vertices.
\begin{fact}
\label{fact:colourcyclebound}
Let $G = (V, E_1, \dots, E_k)$ be an edge-coloured graph. Suppose that $\cdeg(v) \geq k'$ for all $v \in V$ and $G$ does not contain a unique colour cycle. Then, $\abs{V} \geq 2^{k'}$, or equivalently $k' \leq \log_2 \abs{V}$.
\end{fact}
\begin{proof}
The proof follows by induction on $k'$. The base case of $k' = 0$ is trivial. Suppose $k' \geq 1$. Because $k' \geq 1$, there exists $i \in [k]$ such that $E_i$ is nonempty. Let $\{v,v'\} \in E_i$ be an arbitrary edge with colour $i$. Observe that removing all edges in $E_i$ from $G$ must disconnect $v$ and $v'$, as otherwise there is a path from $v$ to $v'$ that uses colours other than $i$, and then adding in the last edge $\{v,v'\} \in E_i$ produces a unique colour cycle.

Deleting all edges of colour $i$ thus splits the graph into two disconnected pieces, and can only decrease the colour degree of any vertex by $1$. By induction, each piece has at least $2^{k'-1}$ vertices, and thus there are at least $2^{k'}$ vertices in the original graph.
\end{proof}

\begin{definition}[Directed graphs]
\label{def:digraph}
A directed graph $D$ is a collection $(V,F)$ where $V$ is the set of vertices and $F \subseteq V \times V$ is a multiset of directed edges. We let $\din(v)$ denote the in-degree of $v$ and $\dout(v)$ denote the out-degree of $v$. For sets $S, T \subseteq V$, we let $e^D(S,T)$ be the number of directed edges of the form $(v,v')$  where $v \in S$ and $v' \in T$. When $D$ is clear from context, we will omit the subscript and simply write $e(S,T)$.

We say that a directed graph $D$ is \emph{Eulerian} if $\din(v) = \dout(v)$ for all $v \in V$.
\end{definition}

The following result, due to Huang, Ma, Shapira, Sudakov and Yuster \cite{HMSSY13}, shows that any dense Eulerian directed graph must have a short cycle.
\begin{fact}[Corollary 1.3 in~\cite{HMSSY13}]
\label{fact:digraphcycle}
Let $D=(V,F)$ be an Eulerian directed graph with no parallel edges. Then $D$ has a directed cycle of length at most $6 \abs{V}^2/\abs{F}$. 
\end{fact}

We will need the following definition of edge expansion, which makes sense for dense directed graphs.
\begin{definition}[Cut-dense directed graphs]
\label{def:denseexpansion}
Let $D = (V,F)$ be a directed graph. We say that $D$ is $\lambda$-cut-dense if for every $S \subseteq V$, it holds that
\begin{equation*}
e(S, V\setminus S) \geq \lambda \abs{S} \abs{V \setminus S}.
\end{equation*} 
\end{definition}

Finally, we will use Cheeger's inequality.
\begin{fact}[Cheeger's inequality]
\label{fact:cheeger}
Let $G = (V,E)$ be an undirected $d$-regular graph, and let $A_G$ be its adjacency matrix. For $S \subseteq V$ with $1 \leq \abs{S} \leq \abs{V}/2$, define
\begin{equation*}
\phi(S) \defeq \frac{e(S, V \setminus S)}{d \abs{S}} \mcom
\end{equation*}
and let $\phi(G) = \min_{1 \leq \abs{S} \leq \abs{V}/2} \phi(S)$. Then, $\lambda_2 \geq \phi(G)^2/2$, where $\lambda_2$ is the second smallest eigenvalue of the matrix $\Id - \frac{1}{d} A_G$.
\end{fact}
\section{Proof of \cref{mthm:main}}
\label{sec:main}
In this section, we prove \cref{mthm:main}. Let $\Code \colon \F_2^k \to \F_2^n$ be a linear $(3, \delta, \eps)$-LCC. By \cref{fact:lcccomb} and changing $\delta$ and $n$ by constant factors, we may assume that we have vectors $\bfv_1, \dots, \bfv_n \in \F_2^k$ with $\dim(\vspan\{\bfv_1, \dots, \bfv_n\}) = k$ and $3$-uniform hypergraph matchings $\cH_1, \dots, \cH_n$, each containing at least $\delta n$ hyperedges, such that for every $u \in [n]$ and $\{u_1, u_2, u_3\} \in \cH_u$, it holds that $\bfv_u = \bfv_{u_1} + \bfv_{u_2} + \bfv_{u_3}$. To prove \cref{mthm:main}, we must show that $k = \dim(\vspan\{\bfv_1, \dots, \bfv_n\}) \leq O(\delta^{-14}\log^2 n)$.

The following key technical lemma completes the proof under the assumption that the directed graph obtained from ``flattening'' the hypergraph matchings is a good edge expander.
\begin{restatable}{lemma}{lbfromexpansion}
\label{lem:lbfromexpansion}
Let $\Omega$ and $\Lambda$ be finite sets of size $m$ and $n$, respectively. For each $u \in \Omega$ and each $a \in \Lambda$, let $\bfv_u$ and $\bfv_a$ be elements of an ambient vector space $\cW$ over $\F_2$.

Let $\cH$ denote a collection of tuples of the form $(u,\{a, b\}, u')$ where $u,u' \in \Omega$ and $\{a,b\} \in {\Lambda \choose 2}$, where for every pair $(u,a)\in \Omega\times \Lambda$ there is at most $1$ tuple of the form $(u, \{a, \star\}, \star)$, and for each $(u,\{a,b\},u') \in \cH$, it holds that $\bfv_u + \bfv_a + \bfv_b + \bfv_{u'} = \mathbf{0}$.

Let $D_{\cH}$ be the directed graph with vertex set $\Omega$ and a directed edge $(u,u')$ for every tuple $(u, \{a, b\}, u') \in \cH$.  Suppose that $D_{\cH}$ is an Eulerian directed graph (\cref{def:digraph}) and is $\lambda$-cut-dense (\cref{def:denseexpansion}). Then, 
\begin{equation*}
\dim(\vspan\{\bfv_u\}_{u \in \Omega}) \leq O\left(\frac{n^2}{\lambda^2 m^2} \cdot \log m \cdot \log n\right) \mper
\end{equation*}
\end{restatable}

For technical reasons, \cref{lem:lbfromexpansion} requires that $D_{\cH}$ is Eulerian. However, it is possible to prove a similar statement without this assumption, at the cost of making the dependence on $1/\lambda$ exponential, rather than polynomial, in the dimension bound.

\cref{lem:lbfromexpansion} is stated in the general form that we will need for the proof. However, for intuition it is helpful to consider the following setting. Take $\Omega = \Lambda = [n]$, and let the vectors $\bfv_u$ and $\bfv_a$ in \cref{lem:lbfromexpansion} be the vectors $\bfv_1, \dots, \bfv_n$ from \cref{fact:lcccomb}. Then, take $\cH$ to denote the tuples obtained by, for every $u \in [n]$ and every hyperedge $\{a,b,u'\} \in \cH_u$, picking an arbitrary ``special vertex'' $u'$ in the hyperedge and adding the tuple $(u, \{a, b\}, u')$ to $\cH$. \cref{lem:lbfromexpansion} essentially states that if the directed graph with the arcs $u \to u'$ obtained from the tuples is a good edge expander, then we have $k \leq O_{\delta}(\log^2 n)$.

We note that this implies\footnote{Up to the assumption that $D_{\cH}$ is Eulerian, which is only needed for the better dependence on $\delta$, see above.} the $k \leq O(\log^2 n)$ bound obtained in~\cite{KothariM24} for the case of ``design LCCs''. Indeed, in a design LCC the matchings are \emph{perfect}, which means that if we pick the special vertex $u'$ randomly from each hyperedge, then the directed graph $D_{\cH}$ is essentially a random directed graph with out-degree $n/3$ at each vertex, which is a good expander. 

In the remainder of this section, we will prove \cref{mthm:main} assuming \cref{lem:lbfromexpansion}.
\begin{proof}[Proof of \cref{mthm:main}]
Let $\cW = \vspan\{\bfv_1, \dots, \bfv_n\}$. We will construct a sequence of subspaces $\{\mathbf{0}\} = \cW_0 \subseteq \cW_1 \subseteq \dots \subseteq \cW_t = \cW$ such that $t \leq 8/\delta^2$ and $\dim(\cW_i) - \dim(\cW_{i-1}) \leq O(\delta^{-12} \log^2 n)$. This clearly implies that $k = \dim(\cW) \leq O(\delta^{-14} \log^2 n)$, which proves \cref{mthm:main}.

We construct $\cW_i$ using $\cW_{i-1}$. Given $\cW_{i-1}$, denote by $\bbfv_u$ the image of $\bfv_u$ in the quotient space $\cW/\cW_{i-1}$. Let $V = \{u \in [n]: \bbfv_u \ne \mathbf{0}\}$ be the subset of indices whose vectors remain nonzero. If $V = \emptyset$, then we stop this process and let $t = i-1$, and note that $\cW_{i-1} = \cW$.

Suppose that $V \ne \emptyset$, and let $u \in V$. Recall that we have a hypergraph matching $\cH_u$ with at least $\delta n$ edges such that for every $\{u_1, u_2, u_3\} \in \cH_u$, we have that $\bfv_{u} + \bfv_{u_1} + \bfv_{u_2} + \bfv_{u_3} = \mathbf{0}$. Note that this relation continues to hold in the quotient space $\cW / \cW_{i-1}$, i.e., $\bbfv_{u} + \bbfv_{u_1} + \bbfv_{u_2} + \bbfv_{u_3} = \mathbf{0}$. Since $\bbfv_{u} \ne \mathbf{0}$, it follows that at least one of $\bbfv_{u_1}$, $\bbfv_{u_2}$, or $\bbfv_{u_3}$ must also be nonzero. Let $u' \in V$ be the index of one of these nonzero vectors, chosen arbitrarily.

We now construct a directed graph $D$ with vertex set $V$ as follows. For each $u \in V$ and each $\{u_1, u_2, u_3\} \in \cH_u$, we add a directed edge $(u,u')$ to $D$, where $u'$ is the element chosen from the hyperedge $\{u_1, u_2, u_3\}$ as described above, so that $\bbfv_{u'} \ne \mathbf{0}$ and thus $u' \in V$. Note that each edge $(u,u')$ in $D$ corresponds to a particular hyperedge $\{u_1, u_2, u'\} \in \cH_u$, and since $\cH_u$ is a hypergraph matching, $D$ has no parallel directed edges. Furthermore, since $\cH_u$ has at least $\delta n$ hyperedges, we have that $\min_{v \in V} \dout(v) \geq \delta n$.

The following lemma, which we prove in \cref{sec:euleriansubgraph}, shows that there exists an Eulerian expanding subgraph of $D$ with a large vertex set.
\begin{restatable}{lemma}{euleriansubgraph}
\label{lem:euleriansubgraph}
Let $D = (V,F)$ be a directed graph with no parallel edges, $\abs{V} \leq n$ and $\min_{v \in V} \dout(v) \geq \delta n$. Then there is an Eulerian subgraph $D' = (V', F')$ of $D$ with $\abs{V'} \geq \delta^2 n/8$ such that $D'$ is $\lambda$-cut dense with $\lambda = \delta^4/96$.
\end{restatable}

We now use \cref{lem:euleriansubgraph} to finish the proof.
The directed graph $D$ defined above satisfies the requirements of \cref{lem:euleriansubgraph}. Let $D' = (V',F')$ be the subgraph of $D$ guaranteed by \cref{lem:euleriansubgraph}, and let $\cW_i = \cW_{i-1} + \vspan\{\bfv_{u} : u \in V'\}$. Note that $\dim(\cW_i) - \dim(\cW_{i-1}) = \dim(\cW_i / \cW_{i-1}) = \dim(\vspan\{\bbfv_{u} : u \in V'\})$, where $\bbfv_u$ is the image of $\bfv_u$ in $\cW / \cW_{i-1}$.

We now apply \cref{lem:lbfromexpansion}. We let $\Omega = V'$ and $\Lambda = [n]$, and the vectors associated with elements of $\Omega$ and $\Lambda$ are the $\bbfv_{u}$'s, which lie in the vector space $\cW / \cW_{i-1}$. We let $\cH$ denote the tuples $(u, \{a, b\}, u')$ where $u \in V'$, $\{a,b,u'\} \in \cH_u$, and the hyperedge $\{a,b,u'\}$ ``corresponds to'' an edge that is in $D'$. In other words, for each edge $(u,u') \in F'$, we ``remember'' the hyperedge $\{a,b,u'\} \in \cH_u$ that produced the edge $(u,u')$ in $D'$, and we add the tuple $(u,\{a,b\},u')$ to $\cH$. Notice that we have $u' \in V'$, and $a,b \in [n]$. Furthermore, because $\cH_u$ is a matching, the first requirement on $\cH$ is satisfied, and the second requirement holds because the linear equation holds in $\cW$, and thus also holds in $\cW / \cW_{i-1}$. The directed graph $D_{\cH}$ in the statement of \cref{lem:lbfromexpansion} is simply $D'$, and so it is an Eulerian directed graph that is $\lambda$-cut-dense. By \cref{lem:lbfromexpansion}, we thus have that 
\begin{equation*}
\dim(\cW_i/\cW_{i-1}) = \dim(\vspan\{\bbfv_{u} : u \in V'\}) \leq O\left(\frac{n^2}{\lambda^2 \abs{V'}^2} \cdot \log \abs{V'} \cdot \log n\right) = O\left(\delta^{-12} \log^2 n\right) \mcom
\end{equation*}
since $\abs{V'} \geq \delta^2 n/8$ and $\lambda = \delta^4/96$.

We have thus shown how to construct $\cW_i$ from $\cW_{i-1}$. It remains to argue that the number of iterations is bounded by $8/\delta^2$. This follows because by \cref{lem:euleriansubgraph}, we have that $\abs{V'} \geq \delta^2 n/8$, and since all the vectors $\bfv_{u}$ for $u \in V'$ are in $\cW_i$, this means that $\abs{V}$ decreases by at least $\delta^2 n/8$ with each iteration. So, the total number of iterations is bounded by $8/\delta^2$, which finishes the proof of \cref{mthm:main}.
\end{proof}

\subsection{Proof of \cref{lem:euleriansubgraph}}
\label{sec:euleriansubgraph}
In this subsection, we prove \cref{lem:euleriansubgraph}, restated below.
\euleriansubgraph*
\begin{proof}
Let $d = \ceil{\delta n}$. Note that $\abs{V} \geq d$ since each $v \in V$ has at least $d$ outgoing edges and there are no parallel edges. We will construct $D'$ in three steps.

As the first step, we will find an Eulerian subgraph of $D$ that has a large fraction of the edges. To do this, we remove cycles from $D$ until the resulting graph $D''$ is acyclic, and let $F_1$ be the union of the removed cycles. Clearly, we have that $D_1 = (V,F_1)$ is Eulerian.

We now show that $\abs{F_1} \geq d^2/2$, so that $D_1$ has many edges. Because $D''$ is a directed acyclic graph, we can order its vertices so that each $v_i$ has no outgoing edges to a later vertex in the order. Hence, in $D''$ the $i$-th vertex in the order has at most $i-1$ outgoing edges remaining, or in other words, at least $d - (i-1)$ outgoing edges have been removed from $D$ and added to $D_1$. Summing over the first $d$ vertices in the order implies that at least
\begin{flalign*}
\sum_{i = 1}^d (d - (i-1)) \geq d^2/2 \geq \delta^2 n^2/2
\end{flalign*}
edges have been removed, and so $\abs{F_1} \geq d^2/2$.

As the second step, we will find an Eulerian subgraph $D_2 = (V, F_2)$ of $D_1$ where $F_2$ is the union of \emph{short} directed cycles. We do this by applying \cref{fact:digraphcycle}. Let $\ell = 24/\delta^2$. Starting from the edge set $F_1$, we repeatedly remove a cycle of length $\leq \ell$ from $D_1$ and add it to $F_2$. Note that by \cref{fact:digraphcycle}, such a cycle exists as long as $\abs{F_1} \geq \delta^2 n^2/4$. Hence, $\abs{F_2} \geq \delta^2 n^2/2 - \delta^2 n^2/4 = \delta^2 n^2/4$, and we have that $F_2 = C_1 \sqcup C_2 \sqcup \dots \sqcup C_t$, where each $C_j$ is a cycle of length at most $\ell$.

As the third and final step, we will split the vertex set of $D_2$ into expanding pieces. However, unlike a standard expander decomposition that removes edges, here we will remove \emph{entire cycles} from $F_2$ at a time, using the cycle decomposition of $F_2$ constructed above, as this will ensure that the resulting graph is Eulerian. That is, when we remove an edge $e$, we will also remove the entire cycle $C_j$, where $C_j$ is the unique cycle in the partition $F_2 = C_1 \sqcup C_2 \sqcup \dots \sqcup C_t$ that contains $e$. The fact that the cycles are \emph{short} will crucially ensure that we do not delete too many edges here.

We now describe this process more formally. Let $\lambda = \delta^4/96$. We will cut our directed graph into a number of disjoint pieces. While there exists a split $U = A \sqcup B$ of a current piece $U$ into nonempty sets where $e^{F_2}(A,B) < \lambda \abs{A} \abs{B}$, we split the graph $D_2$ into two parts as follows. We remove all edges from $A$ to $B$, along with all other edges that are in the unique cycles $C_j$ that contain these edges according to $F_2$'s short cycle decomposition. Because each cycle has length at most $\ell = 24/\delta^2$, each step of this process deletes at most $\ell \lambda \abs{A} \abs{B}$ edges. Note that this process also deletes all edges from $B$ to $A$, as each such edge lies in some cycle $C_j$ in the cycle decomposition of $F_2$, and that cycle must also have an edge from $A$ to $B$. We then replace $U$ by the retained directed graphs on $A$ and $B$.

As each (unordered) pair of vertices $\{u,u'\}$ in $V$ is separated at most once during this process, we have that $\sum_{\textrm{splits } A \sqcup B} \abs{A}\abs{B} \leq {n \choose 2}$. The total number of edges deleted is thus at most $\ell \lambda n^2/2 = \delta^2 n^2/8$. So, after this process terminates, we have decomposed $D_2$ into Eulerian $\lambda$-cut-dense pieces where the total number of edges preserved is at least $\delta^2 n^2/8$.

By averaging, there exists some piece $D' = (V',F')$ where $\abs{F'}/\abs{V'} \geq \delta^2 n/8$. Note that as $\abs{F'} \leq \abs{V'}^2$, this implies that $\abs{V'} \geq \delta^2 n/8$, which finishes the proof.
\end{proof}

\section{The Proof Assuming Edge Expansion: Proof of \cref{lem:lbfromexpansion}}
\label{sec:expansion}
In this section, we prove \cref{lem:lbfromexpansion}, restated below.

\lbfromexpansion*

\begin{proof}
Let $k = \dim(\vspan\{\bfv_{u} : u \in \Omega\})$, 
and $\colourset \subseteq \Omega$ be a set of size $k$ where $\{\bfv_{i} : i \in \colourset\}$
are linearly independent.

We now define a multiset $\cT \subseteq \Omega \times \Lambda \times \Lambda \times \Omega$ as follows. For each tuple $(u, \{a, b\}, u') \in \cH$, we add the tuples $(u, a, b, u')$, $(u, b, a, u')$, $(u,a,a,u)$, and $(u,b,b,u)$ to $\cT$. For each pair $(u,a) \in \Omega \times \Lambda$ where no such tuple exists, we add two copies of $(u, a, a, u)$ to $\cT$. Note that there is at most one tuple of the form $(u, \{a, b\}, u') \in \cH$ for a fixed choice of $(u,a)$, by assumption. Therefore, in the multiset $\cT$, for any choice of $(u,a)$, there are exactly $2$ tuples of the form $(u, a, \star, \star)$, and similarly for any choice of $(u,b)$, there are exactly $2$ tuples of the form $(u, \star, b, \star)$.

We call tuples of the form $(u, a, a, u)$ ``fake'', and the tuples originating from $\cH$ ``real''. Note that for any tuple $(u,a,b,u') \in \cT$, real or fake, it holds that $\bfv_{u} + \bfv_a + \bfv_b + \bfv_{u'} = \mathbf{0}$ in $\cW$. Additionally, for any choice of $(u,a)$, at least one of the $2$ tuples of the form $(u, a, \star, \star)$ is fake.

The proof of \cref{lem:lbfromexpansion} revolves around analyzing combinatorial properties of the edge-coloured Kikuchi graph defined below.

\begin{definition}[$r$-chain Kikuchi graph]
\label{def:kikuchi}
Let $r \geq 1$ be an integer. The graph $\cK_r(\colourset, \cT)$ is a bipartite graph with left vertices $V_L = \Lambda^r \times \Omega$ and right vertices $V_R=\Lambda^r$. We let $(\vec{a}, u)$ where $\vec{a} = (a_1, \dots, a_r)$ denote a left vertex in $V_L$, and $\vec{b} = (b_1, \dots, b_r)$ denote a right vertex in $V_R$.

For each sequence $(i, a_1, b_1, u_1, \dots, u_{r-1}, a_r, b_r, u_r)$ where $i \in \colourset$ and $(u_{j-1}, a_j, b_j, u_j) \in \cT$ for all $j \in [r]$ (treating $u_0 = i$), we add an edge $(((a_1, \dots, a_r), u_r), (b_1, \dots, b_r))$ of colour $i$ to $\cK_r(\colourset,\cT)$. These edges are added with multiplicity according to the multiplicities of the tuples in $\cT$.
\end{definition}
Because the colour set $\colourset$ is chosen so that the $k$ vectors $\{\bfv_{i} : i \in \colourset\}$ are linearly independent, it follows that $\cK_r(\colourset,\cT)$ has no odd-coloured or unique colour cycle. Indeed, this is due to the following observation, which holds because $(u_{j-1}, a_j, b_j, u_j) \in \cT$ implies that $\bfv_{u_{j-1}} + \bfv_{a_j} + \bfv_{b_j} + \bfv_{u_j} = \mathbf{0}$.
\begin{observation}
For a vertex $(\vec{a},u) \in V_L$, let $\bfv_{(\vec{a},u)} \defeq \bfv_u + \sum_{j = 1}^r \bfv_{a_j}$, and for a vertex $\vec{b} \in V_R$, let $\bfv_{\vec{b}} \defeq \sum_{j = 1}^r \bfv_{b_j}$. Then, for any edge $((\vec{a},u), \vec{b})$ in $\cK_r(\colourset, \cT)$ of colour $i \in \colourset$, it holds that $\bfv_{(\vec{a},u)} + \bfv_{\vec{b}} = \bfv_i$. 
\end{observation}
Thus, if $\cK_r(\colourset, \cT)$ contains an odd-coloured cycle, then for each edge in the cycle $((\vec{a},u), \vec{b})$, we have $\bfv_{(\vec{a},u)} + \bfv_{\vec{b}} = \bfv_i$. Summing over all edges in the cycle yields $\mathbf{0}$ on the left, while the right hand side contains some $\bfv_i$ with an odd coefficient. The only other terms that can appear on the right are $\bfv_{i'}$ for $i' \in \colourset$ with $i' \ne i$, which are linearly independent from $\bfv_i$, and thus the right hand side cannot be $\mathbf{0}$.

So, it remains to show that $\cK_r(\colourset, \cT)$ must contain an odd-coloured cycle if $k = \abs{\colourset}$ is too large. We do this by controlling the following variance quantity, which captures the expected local variance of the degrees of each colour at a single vertex.
\begin{equation}
\label{eq:localvar2}
\cV \defeq \E_{v \sim V_L \sqcup V_R} \left[\E_{i \sim \colourset} \left[\left(d_i(v) - \frac{d(v)}{k}\right)^2\right]\right] \mper
\end{equation}
Here, $d_i(v)$ is the degree of the vertex $v$ within the $i$-th colour in $\cK_r(\colourset, \cT)$.

The following lemma, which we prove in \cref{sec:vartocycle}, shows that if we can obtain a sufficient upper bound on $\cV$ in \cref{eq:localvar2}, then we are done.

\begin{restatable}{lemma}{vartocycle}
\label{lem:vartocycle}
Let $G = (V, E_1, \dots, E_k)$ be an edge-coloured graph. Suppose that $G$ has no unique colour cycle. If 
\begin{equation*}
\cV \defeq \E_{v \sim V} \left[\E_{i \sim [k]} \left[\left(d_i(v) - \frac{d(v)}{k}\right)^2\right]\right] \leq \left(\frac{\abs{E}}{k \abs{V}}\right)^2 \mcom
\end{equation*}
then $k \leq 60 \log_2 \abs{V}$.
\end{restatable}

The remainder of the proof is therefore dedicated to bounding $\cV$. We do this by connecting $\cV$ to the behavior of a random walk on the directed transition graph $D_{\cT}$, specified below.
\begin{definition}[Transition graph]
\label{def:transitiongraph}
Let $D_{\cT} = (\Omega, F_{\cT})$ be the directed graph with vertex set $\Omega$, and edge set $F_{\cT}$ defined by adding an edge $(u,u')$ for each $(u,a,b,u') \in \cT$ (with multiplicity). We let $Q_{\cT}$ denote the adjacency matrix of $D_{\cT}$, i.e., $Q_{\cT}(u', u)$ is equal to the number of edges $(u,u')$ in the multiset $F_{\cT}$.

We let $G_{\cT} = (\Omega, E_{\cT})$ be the \emph{undirected} graph with vertex set $\Omega$, defined via the adjacency matrix $A_{\cT} \defeq Q_{\cT} + Q^{\top}_{\cT}$.
\end{definition}
The graphs $D_{\cT}$ and $G_{\cT}$ have the following properties, which we will use.
\begin{proposition}[Properties of the transition graph]
\label{prop:transitionprop}
The following hold for $D_{\cT}$ and $G_{\cT}$.
\begin{enumerate}[(1)]
\item The graph $2 D_{\cH}$, i.e., the graph where every edge in $D_{\cH}$ appears with multiplicity $2$, is a subgraph of $D_{\cT}$.
\item The graph $D_{\cT}$ is Eulerian, and every vertex $u \in \Omega$ has $\din(u) = \dout(u) = 2n$. Furthermore, each $u$ has at least $n$ self-loops.
\item The graph $G_{\cT}$ is $4n$-regular.
\item For each $a \in \Lambda$, let $M_a$ be the matrix where $M_a(u',u) = \frac{1}{2}\abs{\{(u,a,\star,u') \in \cT\}}$. Then, $\E_{a \sim \Lambda}[M_a] = \frac{1}{2n} Q_{\cT}$, every column of $M_a$ sums to $1$ and $M_a(u,u)\geq 1/2$ for all $u$.
\item Let $\lambda_2$ denote the second smallest eigenvalue of the Laplacian $\Id - \frac{1}{4n} A_{\cT}$. Then, $\lambda_2 \geq \lambda^2 m^2/8n^2$, where $m = \abs{\Omega}$.
\end{enumerate}
\end{proposition}
The proof of \cref{prop:transitionprop} is straightforward, and is given below.
\begin{proof}
The first property follows by definition of $\cT$ and the graph $D_{\cT}$, because each $(u, \{a,b\}, u') \in \cH$ produces $2$ real tuples, $(u,a,b,u')$ and $(u,b,a,u')$, in $\cT$.

The second property follows because the graph $D_{\cH}$ in the statement of \cref{lem:lbfromexpansion} is Eulerian, and the addition of the ``fake'' tuples to $\cT$ makes $\dout(u) = 2n$ for all $u \in \Omega$. We also have $\din(u) = 2n$, since the addition of the fake tuples preserves the Eulerian property. The fake tuples in $\cT$ provide at least $n$ self-loops at each vertex.

The third property follows from the second because $G_{\cT}$ is obtained by removing the direction of the edges in $D_{\cT}$. Hence, $d^{G_{\cT}}(u) = \din^{D_{\cT}}(u) + \dout^{D_{\cT}}(u) = 4n$.

The fourth property is by definition of the adjacency matrix. The fact that the column sums of $M_a$ are $1$ follows because for every $(u,a)$, there are exactly $2$ tuples of the form $(u,a, \star, \star) \in \cT$, and $M_a(u,u)\geq 1/2$ follows from the existence of at least one fake tuple $(u,a,a,u)$ in $\cT$.

The final property follows from Cheeger's inequality (\cref{fact:cheeger}) and the fact that the graph $D_{\cH}$ is $\lambda$-cut-dense (\cref{def:denseexpansion}). Indeed, for any nonempty $S \subseteq \Omega$, we have $\phi(S) = e^{G_{\cT}}(S, \Omega \setminus S) / 4 n \abs{S}$. Because $2D_{\cH}$ is a subgraph of $D_{\cT}$ and is $\lambda$-cut-dense, we have
\begin{equation*}
e^{G_{\cT}}(S, \Omega \setminus S) \geq 2e^{D_{\cH}}(S, \Omega \setminus S) + 2e^{D_{\cH}}(\Omega \setminus S, S) \geq 4\lambda \abs{S} \abs{\Omega \setminus S} \mper
\end{equation*}
Hence, $\phi(S) \geq \frac{\lambda\abs{\Omega \setminus S}}{n}$, and so $\phi(G_{\cT}) \geq \frac{\lambda m}{2 n}$. By \cref{fact:cheeger}, this implies that $\lambda_2 \geq \lambda^2 m^2/8n^2$.
\end{proof}

With \cref{def:transitiongraph,prop:transitionprop} in hand, we are ready to bound $\cV$. First, we will show that $\cV$ is \emph{equal} to a quantity related to a random walk on the graph $D_{\cT}$. This quantity is meant to capture the difference in distributions obtained by running a random walk for $r$ steps, starting either at the vertex $i \in \Omega$, or another vertex $i' \in \Omega$. Then, we will bound this quantity using Item (5) in \cref{prop:transitionprop}, which says that $\lambda_2$ of the Laplacian of $G_{\cT}$ is somewhat large.

The claim below relates $\cV$ to the random walk quantity, and then the lemma bounds the quantity.
\begin{claim}
\label{claim:vartorandomwalk}
For $a \in \Lambda$, let $M_a$ be the matrix where $M_a(u',u) = \frac{1}{2}\abs{\{(u,a,\star,u') \in \cT\}}$. Then, it holds that
\begin{equation*}
\cV = \frac{2^{2r}}{2(m+1)} \E_{i, i' \in \colourset}\left[ \E_{a_1, \dots, a_r \sim \Lambda}\left[\norm{ M_{a_r} \dots M_{a_1}(e_i - e_{i'})}_2^2\right] \right] \mper
\end{equation*}
\end{claim}
\begin{restatable}{lemma}{mixing}
\label{lem:mixing}
For $r \in \N$, let $M^{(1)}, \dots, M^{(r)} \in \R^{m \times m}$ be independent copies of a random matrix $M$, where with probability $1$, it holds that (1) $M$ is column stochastic: it has nonnegative entries and column sums equal to $1$, and (2) $M \geq \Id/2$ entrywise. Suppose that $\overline{M} = \E[M]$ is doubly stochastic: all entries are nonnegative and all row and column sums are $1$, and suppose that the second largest eigenvalue of $\frac{1}{2}(\overline{M} + \overline{M}^{\top})$ is at most $1 - \lambda_2$. Then,
\begin{equation*}
\max_{u,u' \in [m]} \E_{M^{(1)}, \dots, M^{(r)}}\left[ \norm{M^{(r)} \dots M^{(1)} (e_u - e_{u'})}_2^2\right] \leq 2 m(1 - \lambda_2)^r \mper
\end{equation*}
\end{restatable}

The proof of \cref{claim:vartorandomwalk} is simple, and we give it below. We postpone the proof of \cref{lem:mixing} to \cref{sec:mixing}.
\begin{proof}[Proof of \cref{claim:vartorandomwalk}]
First, we observe that each right vertex $\vec{b} \in V_R$ contributes $0$ to $\cV$. Indeed, we have shown that for each pair $(u,b)$, there are exactly $2$ tuples of the form $(u, \star, b, \star) \in \cT$, and this implies that $d_i(\vec{b}) = 2^r$ for all $i$ and all $\vec{b}$. Since $\abs{V_L} = \abs{\Omega} \abs{V_R} = m \abs{V_R}$, we have that
\begin{equation*}
\cV = \frac{m}{m+1} \E_{v \sim V_L} \left[\E_{i \sim \colourset} \left[\left(d_i(v) - \frac{d(v)}{k}\right)^2\right]\right] \mper
\end{equation*}
We also have that for any $v \in V_L$,
\begin{equation*}
\E_{i \sim \colourset} \left[\left(d_i(v) - \frac{d(v)}{k}\right)^2\right] = \frac{1}{2}\E_{i, i' \sim \colourset}\left[\left(d_i(v) - d_{i'}(v)\right)^2\right] \mper
\end{equation*}
Next, we observe that for any $(\vec{a}, u) \in V_L$, we have that $d_i(\vec{a}, u) = e_{u}^{\top} 2^r M_{a_r} \dots M_{a_1} e_i$. This follows because there are exactly $2$ tuples of the form $(u, a, \star, \star) \in \cT$, and thus the $u$-th entry of $2^r M_{a_r} \dots M_{a_1} e_i$ exactly counts the number of $r$-chains of the form $(i, a_1, \star, u_1, a_2, \star, \dots, u_{r-1}, a_r, \star, u)$, which is exactly $d_i(\vec{a},u)$. Hence,
\begin{flalign*}
&\cV = \frac{m}{2(m+1)} \E_{v \sim V_L}\left[ \E_{i, i' \sim \colourset}\left[\left(d_i(v) - d_{i'}(v)\right)^2\right]\right] \\
&= \frac{m}{2(m+1)}  \E_{i, i' \sim \colourset}\left[\E_{v \sim V_L}\left[\left(d_i(v) - d_{i'}(v)\right)^2\right]\right] \\
&= \frac{m}{2(m+1)}  \E_{i, i' \sim \colourset}\left[\E_{a_1, \dots, a_r \sim \Lambda, u \sim \Omega}\left[\left(d_i(\vec{a},u) - d_{i'}(\vec{a},u)\right)^2\right]\right] \\
&=\frac{m}{2(m+1)} \cdot 2^{2r} \frac{1}{m} \cdot \E_{i, i' \sim \colourset}\left[ \E_{a_1, \dots, a_r \sim \Lambda} \left[ \norm{M_{a_r} \dots M_{a_1} (e_i - e_{i'})}_2^2\right] \right] \mcom
\end{flalign*}
which finishes the proof.
\end{proof}

We are now ready to finish the proof of \cref{lem:lbfromexpansion}. By \cref{claim:vartorandomwalk}, we have 
\begin{equation*}
\cV = \frac{2^{2r}}{2(m+1)} \E_{i, i' \in \colourset}\left[ \E_{a_1, \dots, a_r \sim \Lambda}\left[\norm{ M_{a_r} \dots M_{a_1}(e_i - e_{i'})}_2^2\right] \right] \mper
\end{equation*}
We now apply \cref{lem:mixing} to bound $\E_{a_1, \dots, a_r \sim \Lambda}\left[\norm{ M_{a_r} \dots M_{a_1}(e_i - e_{i'})}_2^2\right]$. Indeed, we have that each $M_{a_j}$ is an independent copy of the same random matrix, and each $M_{a_j}$ has column sums equal to $1$ by Item (4) in \cref{prop:transitionprop}. Each $M_{a_j}$ is furthermore entrywise at least $\Id/2$, as all entries are nonnegative, and the diagonal entries are at least $1/2$ because of Item (4) in \cref{prop:transitionprop}. The mean $\overline{M}$ is equal to $\E_{a \sim \Lambda}[M_a]$, which by Item (4) in \cref{prop:transitionprop} is equal to $\frac{1}{2n} Q_{\cT}$. By Item (2) in \cref{prop:transitionprop}, $\frac{1}{2n} Q_{\cT}$ is doubly stochastic. By Item (5) in \cref{prop:transitionprop}, the matrix $\frac{1}{2}(\overline{M} + \overline{M}^{\top}) = \frac{1}{4n} A_{\cT}$ has second largest eigenvalue at most $1 - \lambda_2$ for $\lambda_2 \geq \lambda^2 m^2/8n^2$. Thus, applying \cref{lem:mixing}, we conclude that
\begin{equation*}
\cV = \frac{2^{2r}}{2(m+1)} \E_{i, i' \in \colourset}\left[ \E_{a_1, \dots, a_r \sim \Lambda}\left[\norm{ M_{a_r} \dots M_{a_1}(e_i - e_{i'})}_2^2\right] \right] \leq \frac{2^{2r}}{2(m+1)} \cdot 2m \left(1 - \frac{\lambda^2 m^2}{8n^2}\right)^r \mper
\end{equation*}

Finally, in the Kikuchi graph $\cK_r(\colourset, \cT)$, we have $\abs{V} = \abs{V_L} + \abs{V_R} = n^r (m+1)$ and $\abs{E} = \abs{V_R} \cdot k \cdot 2^r = k (2n)^r$. Hence, $\abs{E}/k\abs{V} = 2^r/(m+1)$. Thus, taking $r = c \frac{n^2}{\lambda^2 m^2} \cdot \log_2 m$ for a sufficiently large constant $c$, we get that $\cV \leq \left(\abs{E}/k\abs{V}\right)^2$, and by \cref{lem:vartocycle} we conclude that $k \leq 60\log_2 \abs{V} = O(r \log n + \log m) = O(\frac{n^2}{\lambda^2 m^2} \cdot \log m \cdot \log n)$. This finishes the proof up to the proofs of \cref{lem:vartocycle} and \cref{lem:mixing}, which we show in the subsequent subsections.
\end{proof}

\subsection{Proof of \cref{lem:vartocycle}}
\label{sec:vartocycle}
In this subsection, we prove \cref{lem:vartocycle}, restated below.
\vartocycle*

\begin{proof}
Before we give the detailed proof, we will explain the ideas behind the calculations, which are relatively simple. Using the bound on $\cV$, one can show that there are few edges of colour $i$ that touch vertices $v$ where $d_i(v)$ is much larger than $d(v)/k$. After we remove these bad edges, we still have many edges left, as well as a good upper bound on the degrees of individual colours at every vertex, i.e., $d_i(v)$ for all $i$ and $v$. We then iteratively delete vertices whose remaining degree is small relative to its degree $d(v)$ in the initial graph. This gives us a nonempty subgraph $G'$ of $G$ where every vertex has $d^{G'}_i(v) \leq O(d^{G'}(v)/k)$ for all colours $i$, and thus each vertex has colour degree $\cdeg(v) \geq \Omega(k)$. Applying \cref{fact:colourcyclebound}, we thus obtain an upper bound on $k$.

Let us now turn to the details.
Call a pair $(i,v)$ ``bad'' if $d_i(v) > 10d(v)/k$, or in other words, if colour $i$ is overrepresented at $v$.
Delete every edge $(v,v')$ of colour $i$ if either $(i,v)$ or $(i,v')$ is bad. To bound the number of deleted edges, fix a bad
pair $(i,v)$. Because $(i,v)$ is bad, we have
\begin{equation*}
 d_i(v)-\frac{d(v)}k>\frac{9}{10}d_i(v).
\end{equation*}
Hence, writing $B$ for the number of bad $(i,v)$, we have
\begin{equation*}
 \left(\frac{9}{10}\right)^2\left(\sum_{(i,v)\text{ bad}}d_i(v)\right)^2 \leq \left(\sum_{(i,v)\text{ bad}}\left(d_i(v)-\frac{d(v)}{k}\right)\right)^2
 \leq B
       \sum_v\sum_{i=1}^k
       \left(d_i(v)-\frac{d(v)}k\right)^2
 =B\cdot \mathcal{V} \cdot k \abs{V} \mcom
\end{equation*}
where in the second step we used the Cauchy--Schwarz inequality. Clearly, each vertex $v$ is in at most $k/10$ bad pairs, so $B\leq \abs{V} k/10$. Hence, we obtain $\left(\frac{9}{10}\right)^2(\sum_{(i,v)\text{ bad}}d_i(v))^2\leq \frac{\abs{V}k}{10}\cdot \frac{\abs{E}^2}{k\abs{V}}=\frac{\abs{E}^2}{10}$, so the number of deleted edges is at most $\sum_{(i,v)\text{ bad}}d_i(v)\leq \abs{E}/2$.

In the remaining graph, iteratively delete any vertex $v$ whose current degree is at most $d(v)/6$, where $d(v)$ is the degree of $v$ in the original graph $G$.
This procedure deletes at most $\sum_v d(v)/6=\abs{E}/3$
edges. Thus, the final set of edges $E'$ has size at least
$\abs{E}-\abs{E}/2-\abs{E}/3=\abs{E}/6>0$ edges. Let $G' = (V',E')$ be the final graph after removing all vertices that are not adjacent to any edge in $E'$. 

Every remaining vertex $v \in V'$ satisfies
\begin{equation*}
 d^{G'}(v)\geq\frac{d(v)}{6},
 \qquad
 d_i^{G'}(v)\leq\frac{10d(v)}k
 \quad\text{for every colour }i.
\end{equation*}
Let the \emph{colour degree} $\cdeg(v)$ of a vertex in a graph
be the number of distinct colours incident at that vertex. By the above inequalities, each vertex $v$ in $G'$ has $\cdeg(v) \geq k/60$.

Applying \cref{fact:colourcyclebound} to $G'$, we conclude that $k/60 \leq \log_2 \abs{V'} \leq \log_2 \abs{V}$, and so $k \leq 60 \log_2 \abs{V}$.
\end{proof}

\subsection{Proof of \cref{lem:mixing}}
\label{sec:mixing}
In this subsection, we prove \cref{lem:mixing}, restated below.
\mixing*
\begin{proof}
Let $X \in \R^{m \times m}$ be an arbitrary real matrix. For each $i \in [m]$, let $X_{i, \star}$ denote the $i$-th row of $X$, and let $X_{\star, j}$ denote the $j$-th column. Let $\bar{X} \defeq \frac{1}{m} \sum_{j} X_{\star, j}$ denote the average of the columns. We define the following quantity, which intuitively captures how similar the columns of $X$ are: \begin{equation*}
\Ener(X) \defeq \sum_{j = 1}^m \norm{X_{\star,j} - \bar{X}}_2^2 \mper
\end{equation*}
We will argue that for any matrix $X \in \R^{m \times m}$, it holds that $\E_{M}[\Ener(X M)] \leq (1 - \lambda_2) \Ener(X)$. Intuitively, this means that multiplying by $M$ on the right makes the columns of $X$ become more similar. 

This bound immediately implies \cref{lem:mixing} via the following argument. We have
\begin{equation*}
\E_{M^{(r)}, \dots, M^{(2)}}[\E_{M^{(1)}}[\Ener(M^{(r)} \dots M^{(2)} M^{(1)})]] \leq (1 - \lambda_2) \E_{M^{(r)}, \dots, M^{(2)}}[\Ener(M^{(r)} \dots M^{(2)})] \leq (1 - \lambda_2)^r \Ener(\Id) \leq m (1 - \lambda_2)^r\mcom
\end{equation*}
by induction on $r$, since $\Ener(\Id) \leq m$.
Letting $Y = M^{(r)} \dots M^{(1)}$, we have that for any distinct $u, u' \in [m]$,
\begin{equation*}
\norm{Y (e_u - e_{u'})}_2^2 = \norm{Y_{\star, u} - Y_{\star, u'}}_2^2 \leq 2 \norm{Y_{\star, u} - \overline{Y}}_2^2 + 2 \norm{Y_{\star, u'} - \overline{Y}}_2^2 \leq 2 \Ener(Y) \mper
\end{equation*}
Hence, 
\begin{equation*}
\max_{u, u' \in [m]} \E_{M^{(1)}, \dots, M^{(r)}}[\norm{M^{(r)} \dots M^{(1)} (e_u - e_{u'})}_2^2] \leq 2 \E_{M^{(1)}, \dots, M^{(r)}}[\Ener(M^{(r)} \dots M^{(1)})] \leq 2 m (1 - \lambda_2)^r \mcom
\end{equation*}
as required.

It thus remains to show that for any $X \in \R^{m \times m}$, it holds that $\E_{M}[\Ener(X M)] \leq (1 - \lambda_2) \Ener(X)$. 

First, we observe that because $M$ is column stochastic with probability $1$, we can without loss of generality assume that $\bar{X} = \mathbf{0}$. In other words, we may replace $X$ with $X - \bar{X} \mathbf{1}^{\top}$. Indeed, by definition we have that $\Ener(X) = \Ener(X - \bar{X} \mathbf{1}^{\top})$. Since $M$ is column stochastic with probability $1$, it holds that $\mathbf{1}^{\top} M = \mathbf{1}^{\top}$, and therefore $\Ener((X - \bar{X} \mathbf{1}^{\top}) M) = \Ener(XM  - \bar{X} \mathbf{1}^{\top}) = \Ener(X M)$. Note that if $\bar{X} = \mathbf{0}$, then we have $\Ener(X) = \sum_{j = 1}^m \norm{X_{\star,j}}_2^2 = \norm{X}_F^2 = \sum_{i = 1}^m \norm{X_{i,\star}}_2^2$.

Next, we observe that for any $y_1, \dots, y_m \in \R^m$, we have that $\sum_{j = 1}^m \norm{y_j - \bar{y}}_2^2 \leq \sum_{j = 1}^m \norm{y_j}_2^2$ where $\bar{y} = \frac{1}{m} \sum_{j = 1}^m y_j$. Therefore, we have
\begin{equation*}
\E_M[\Ener(X M)] \leq \E_M\left[\sum_{j = 1}^m \norm{ (X M)_{\star, j}}_2^2 \right] \mper
\end{equation*}
Now, by assumption we have that $M$ is entrywise at least $\Id/2$ with probability $1$. Therefore, letting $R = 2 M - \Id$, we have that $R$ is also a column stochastic matrix with probability $1$. Since $\overline{M} = \E[M]$ is a doubly stochastic matrix, it follows that the mean $\overline{R} = \E_M[R]$ is also a doubly stochastic matrix. Using $M = \frac{1}{2}(\Id + R)$, we have
\begin{equation*}
\E_M[\Ener(X M)]  \leq \frac{1}{4}\E_M\left[\sum_{j = 1}^m \norm{ X_{\star, j} + (X R)_{\star, j}}_2^2 \right] = \frac{1}{4} \E_M \left[ \sum_{j = 1}^m \left( \norm{X_{\star, j}}_2^2 + \norm{(XR)_{\star, j}}_2^2 + 2\ip{X_{\star, j}, (XR)_{\star, j}}\right) \right] \mper
\end{equation*}
We now bound the terms in the above sum. First, we have $\sum_{j = 1}^m \norm{X_{\star, j}}_2^2 = \Ener(X)$, as we have assumed that $\bar{X} = \mathbf{0}$. Second, because $R$ is column stochastic with probability $1$ and $\overline{R} = \E_M[R]$ is doubly stochastic, we have
\begin{flalign*}
\E_M\left[\sum_{j = 1}^m\norm{(XR)_{\star, j}}_2^2\right] 
&=\E_M\left[ \sum_{j = 1}^m \Norm{\sum_{i = 1}^m R_{ij} X_{\star, i}}_2^2\right] \leq \E_M\left[ \sum_{j = 1}^m \sum_{i = 1}^m R_{ij}\cdot\norm{ X_{\star, i}}_2^2\right] \\
&=\sum_{i = 1}^m \norm{ X_{\star, i}}_2^2 \left(\sum_{j = 1}^m \E_M[R_{ij}]\right) = \sum_{i = 1}^m \norm{ X_{\star, i}}_2^2 = \Ener(X) \mcom
\end{flalign*}
where the inequality is by Jensen's inequality, since $R_{ij} \geq 0$ for all $i,j$ and $\sum_{i = 1}^m R_{ij} = 1$ for all $j$, and the penultimate equality is because $\overline{R} = \E_M[R]$ is doubly stochastic. This bounds the second term in the sum.

We now bound the third and final term. We have
\begin{flalign*}
&\sum_{j = 1}^m \ip{X_{\star,j}, (XR)_{\star,j}} = \sum_{j = 1}^m \ip{X_{\star,j}, \sum_{i = 1}^m R_{ij} X_{\star, i}} = \sum_{j = 1}^m \sum_{k = 1}^m \sum_{i = 1}^m R_{ij} X_{kj} X_{ki}\\
&= \sum_{k = 1}^m  \left(\sum_{i = 1}^m \sum_{j = 1}^m R_{ij} X_{kj} X_{ki}\right) = \sum_{k = 1}^m  (X_{k, \star})^{\top} R X_{k,\star} = \sum_{k = 1}^m  \frac{1}{2}(X_{k, \star})^{\top} \left(R + R^{\top}\right) X_{k,\star} \mcom
\end{flalign*}
where we view $X_{k,\star}$ as a column vector and the last equality uses that $(X_{k, \star})^{\top} R X_{k,\star}$ is a scalar. Now, because $\bar{X} = \mathbf{0}$, it follows that each row $X_{k,\star}$ is orthogonal to $\mathbf{1}$. Since $\overline{R} = \E_M[R]$ is a doubly stochastic matrix, the top eigenvector of $\frac{1}{2}(\overline{R} + \overline{R}^\top)$ is $\mathbf{1}$ with eigenvalue $1$. Furthermore, because $\frac{1}{2}(\overline{M} + \overline{M}^{\top})$ has second eigenvalue at most $1 - \lambda_2$, it follows that $\frac{1}{2}(\overline{R} + \overline{R}^\top)$ has second eigenvalue at most $2(1 - \lambda_2) - 1 = 1 - 2 \lambda_2$.

Therefore, by taking expectation over $M$, we have
\begin{flalign*}
&\E_M\left[\sum_{j = 1}^m \ip{X_{\star,j}, (XR)_{\star,j}}\right] = \sum_{k = 1}^m  \frac{1}{2}(X_{k, \star})^{\top} \left(\overline{R} + \overline{R}^{\top}\right) X_{k,\star} \leq \sum_{k = 1}^m \norm{X_{k, \star}}_2^2 (1 - 2 \lambda_2) = (1 - 2 \lambda_2) \Ener(X) \mcom
\end{flalign*}
where the inequality uses the second eigenvalue bound on $\frac{1}{2}(\overline{R} + \overline{R}^{\top})$ along with the fact that each row $X_{k, \star}$ is orthogonal to $\mathbf{1}$, and the last equality uses that $\Ener(X) = \norm{X}_F^2$ since $\bar{X} = \mathbf{0}$.

Combining all three bounds, we have shown that
\begin{equation*}
\E_M[\Ener(X M)] \leq \frac{1}{4} \left( \Ener(X) + \Ener(X) + 2 (1 - 2 \lambda_2) \Ener(X)\right) = (1 - \lambda_2)\Ener(X) \mcom
\end{equation*}
as required, which finishes the proof.
\end{proof}
\section*{Statement of AI Use}
This statement is intended solely to describe the role of AI in the research process and should not be read as an endorsement of such systems or of the companies that produce them. In the production of this paper, we have used AI models for the following tasks.
\begin{flushleft}
\begin{tabular}{lc}
\textbf{Activity} & \textbf{AI used} \\
Asking the research question & [No] \\
Coming up with the approach & [No] \\
Proof development & [Yes] \\
Writing and exposition & [No] \\
Checking for bugs & [Yes] \\
Other supporting tasks & [Yes] \\
\end{tabular}
\end{flushleft}

\parhead{Details.} The question of proving a $k \leq O_{\delta}(\log^2 n)$ bound for binary linear $3$-LCCs is a known problem left open by previous work~\cite{KothariM23,Yankovitz24,AlrabiahG24,KothariM24}. We began working on this question in September 2024, and we developed the high-level approach used in \cref{sec:expansion} ourselves by January 2025. Specifically, we set out to show that the Kikuchi graph (\cref{def:kikuchi}) has an odd-coloured cycle by bounding the expected local variance of degrees $\cV$ (\cref{eq:localvar2}), and we had proved \cref{lem:vartocycle} ourselves, which shows that bounding $\cV$ is sufficient to finish the proof. We additionally had some more informal intuition that if we could prove that the directed graph $D_{\cH}$ (defined in \cref{lem:lbfromexpansion}) was an edge expander, then we could show that $\cV \leq (\abs{E}/k\abs{V})^2$. However, we did not write down a full formal proof of this, as there was an issue that a random walk on the directed graph $D_{\cH}$ could be periodic, which would prevent it from mixing.

On Oct.\ 5, we gave GPT-6 Astra our notes, which included the main approach of using odd-coloured cycles in Kikuchi graphs with chain-derived constraints, our proof of \cref{lem:vartocycle}, and the conclusion that showing $\cV \leq (\abs{E}/k\abs{V})^2$ is sufficient to prove \cref{mthm:main}. From this input, GPT-6 Astra completed the proof. Over the course of several more interactions, we substantially simplified Astra's completion of our proof approach, obtaining the proof that we have presented in this paper.

From our perspective, the missing key ideas that Astra provided are as follows. Firstly, in \cref{sec:expansion}, while nearly all of the major ideas are ours, we crucially did not see the simple idea of adding the ``fake tuples'' to $\cT$. The addition of fake tuples makes the corresponding random walk \emph{lazy}, and this deals with the periodicity issues we were encountering in a very clean manner. It also makes it so any walk can always be continued, even if, say, $\cH_u$ does not have a hyperedge containing some $a \in [n]$. Astra also provided a proof of \cref{claim:vartorandomwalk}, and while it is a very nice observation that makes the proof particularly clean, in our view it is not an essential component. In fact, in the initial proof Astra provided, the proof contained a different, more technical statement.

Secondly, we did not see how to execute the decomposition done in \cref{sec:main}, and this was entirely provided by Astra. In our view, the decomposition has a very intriguing strategy of working in the \emph{quotient spaces} $\cW /\cW_{i-1}$ and retaining only indices whose vectors remain nonzero, which preserves large out-degree in the remaining vertex set. This was a technical issue that we did not see how to overcome when we thought about using edge-expander decompositions ourselves.
The decomposition also has some additional technical components that ensure the resulting directed graph $D_{\cH}$ produced is Eulerian. However, as we have noted in \cref{sec:main}, this is not essential as one can prove a version of \cref{lem:lbfromexpansion} without the Eulerian assumption, albeit with an exponential dependence on $1/\lambda$.

Finally, we note that the proof of \cref{fact:colourcyclebound} presented here is due to Astra. However, we had proven essentially the same statement ourselves, with just a slightly weaker bound of $k' \leq O(\log  \abs{V})$, which works equally well for its application. We have chosen to use the AI proof in this paper because it is cleaner.

\bibliographystyle{alpha}
\bibliography{references.bib}

\newcommand{\etalchar}[1]{$^{#1}$}
\begin{thebibliography}{AGKM23}

\bibitem[ABS{\etalchar{+}}25]{AlonBSZZ25}
Noga Alon, Matija Buci{\'c}, Lisa Sauermann, Dmitrii Zakharov, and Or~Zamir.
\newblock Essentially tight bounds for rainbow cycles in proper
  edge-colourings.
\newblock {\em Proceedings of the London Mathematical Society}, 130(4):e70044,
  2025.

\bibitem[AG24]{AlrabiahG24}
Omar Alrabiah and Venkatesan Guruswami.
\newblock Near-tight bounds for 3-query locally correctable binary linear codes
  via rainbow cycles.
\newblock In {\em 65th {IEEE} Annual Symposium on Foundations of Computer
  Science, {FOCS} 2024, Chicago, IL, USA, October 27-30, 2024}. {IEEE}, 2024.

\bibitem[AGKM23]{AlrabiahGKM23}
Omar Alrabiah, Venkatesan Guruswami, Pravesh~K. Kothari, and Peter Manohar.
\newblock A near-cubic lower bound for 3-query locally decodable codes from
  semirandom {CSP} refutation.
\newblock In {\em Proceedings of the 55th Annual {ACM} Symposium on Theory of
  Computing, {STOC} 2023, Orlando, FL, USA, June 20-23, 2023}, pages
  1438--1448. {ACM}, 2023.

\bibitem[ALM{\etalchar{+}}98]{AroraLMSS98}
Sanjeev Arora, Carsten Lund, Rajeev Motwani, Madhu Sudan, and Mario Szegedy.
\newblock Proof verification and the hardness of approximation problems.
\newblock {\em Journal of the ACM (JACM)}, 45(3):501--555, 1998.

\bibitem[AS98]{AroraS98}
Sanjeev Arora and Shmuel Safra.
\newblock Probabilistic checking of proofs: A new characterization of np.
\newblock {\em Journal of the ACM (JACM)}, 45(1):70--122, 1998.

\bibitem[BCG20]{BhattacharyyaCG20}
Arnab Bhattacharyya, L~Sunil Chandran, and Suprovat Ghoshal.
\newblock Combinatorial lower bounds for 3-query ldcs.
\newblock In {\em 11th Innovations in Theoretical Computer Science Conference
  (ITCS 2020)}, volume 151, page~85. Schloss Dagstuhl--Leibniz-Zentrum fuer
  Informatik, 2020.

\bibitem[BDL13]{BhowmickDL13}
Abhishek Bhowmick, Zeev Dvir, and Shachar Lovett.
\newblock New bounds for matching vector families.
\newblock In {\em Symposium on Theory of Computing Conference, STOC'13, Palo
  Alto, CA, USA, June 1-4, 2013}, pages 823--832. {ACM}, 2013.

\bibitem[BFNW93]{BabaiFNW93}
L{\'{a}}szl{\'{o}} Babai, Lance Fortnow, Noam Nisan, and Avi Wigderson.
\newblock {BPP} has subexponential time simulations unless {EXPTIME} has
  publishable proofs.
\newblock {\em Comput. Complex.}, 3:307--318, 1993.

\bibitem[BGT17]{BhattacharyyaGT17}
Arnab Bhattacharyya, Sivakanth Gopi, and Avishay Tal.
\newblock Lower bounds for 2-query lccs over large alphabet.
\newblock In {\em Approximation, Randomization, and Combinatorial Optimization.
  Algorithms and Techniques (APPROX/RANDOM 2017)}. Schloss
  Dagstuhl-Leibniz-Zentrum fuer Informatik, 2017.

\bibitem[BHKL25]{BasuHKL25}
Arpon Basu, Jun{-}Ting Hsieh, Pravesh~K. Kothari, and Andrew~D. Lin.
\newblock Improved lower bounds for all odd-query locally decodable codes.
\newblock In {\em 66th {IEEE} Annual Symposium on Foundations of Computer
  Science, {FOCS} 2025, Sydney, Australia, December 14-17, 2025}, pages
  1262--1285. {IEEE}, 2025.

\bibitem[BIW10]{BarkolIW10}
Omer Barkol, Yuval Ishai, and Enav Weinreb.
\newblock On locally decodable codes, self-correctable codes, and t-private
  pir.
\newblock {\em Algorithmica}, 58(4):831--859, 2010.

\bibitem[BK95]{BlumK95}
Manuel Blum and Sampath Kannan.
\newblock Designing programs that check their work.
\newblock {\em Journal of the ACM (JACM)}, 42(1):269--291, 1995.

\bibitem[BLR93]{BlumLR93}
Manuel Blum, Michael Luby, and Ronitt Rubinfeld.
\newblock Self-testing/correcting with applications to numerical problems.
\newblock {\em Journal of computer and system sciences}, 47(3):549--595, 1993.

\bibitem[DLS13]{DLS13}
Shagnik Das, Choongbum Lee, and Benny Sudakov.
\newblock Rainbow {Tur{\'a}n} problem for even cycles.
\newblock {\em European Journal of Combinatorics}, 34(5):905--915, 2013.

\bibitem[Dvi10]{Dvir10}
Zeev Dvir.
\newblock On matrix rigidity and locally self-correctable codes.
\newblock In {\em Proceedings of the 25th Annual {IEEE} Conference on
  Computational Complexity, {CCC} 2010, Cambridge, Massachusetts, USA, June
  9-12, 2010}, pages 291--298. {IEEE} Computer Society, 2010.

\bibitem[Dvi12]{Dvir12}
Zeev Dvir.
\newblock Incidence theorems and their applications.
\newblock {\em CoRR}, abs/1208.5073, 2012.

\bibitem[Dvi16]{Dvir16}
Zeev Dvir.
\newblock Lecture notes on linear locally decodable codes.
\newblock \url{https://www.cs.princeton.edu/~zdvir/LDCnotes/LDC8.pdf}, Fall
  2016.

\bibitem[GKM22]{GuruswamiKM22}
Venkatesan Guruswami, Pravesh~K. Kothari, and Peter Manohar.
\newblock {Algorithms and certificates for Boolean {CSP} refutation: smoothed
  is no harder than random}.
\newblock In {\em {STOC} '22: 54th Annual {ACM} {SIGACT} Symposium on Theory of
  Computing, Rome, Italy, June 20 - 24, 2022}, pages 678--689. {ACM}, 2022.

\bibitem[GKO{\etalchar{+}}17]{GopiKORS17}
Sivakanth Gopi, Swastik Kopparty, {Rafael Mendes de} Oliveira, Noga Ron{-}Zewi,
  and Shubhangi Saraf.
\newblock Locally testable and locally correctable codes approaching the
  gilbert-varshamov bound.
\newblock In {\em Proceedings of the Twenty-Eighth Annual {ACM-SIAM} Symposium
  on Discrete Algorithms, {SODA} 2017, Barcelona, Spain, Hotel Porta Fira,
  January 16-19}, pages 2073--2091. {SIAM}, 2017.

\bibitem[GKST02]{GoldreichKST02}
Oded Goldreich, Howard~J. Karloff, Leonard~J. Schulman, and Luca Trevisan.
\newblock Lower bounds for linear locally decodable codes and private
  information retrieval.
\newblock In {\em Proceedings of the 17th Annual {IEEE} Conference on
  Computational Complexity, Montr{\'{e}}al, Qu{\'{e}}bec, Canada, May 21-24,
  2002}, pages 175--183. {IEEE} Computer Society, 2002.

\bibitem[HKM23]{HsiehKM23}
Jun{-}Ting Hsieh, Pravesh~K. Kothari, and Sidhanth Mohanty.
\newblock {A simple and sharper proof of the hypergraph Moore bound}.
\newblock In {\em Proceedings of the 2023 {ACM-SIAM} Symposium on Discrete
  Algorithms, {SODA} 2023, Florence, Italy, January 22-25, 2023}, pages
  2324--2344. {SIAM}, 2023.

\bibitem[HMS{\etalchar{+}}13]{HMSSY13}
Hao Huang, Jie Ma, Asaf Shapira, Benny Sudakov, and Raphael Yuster.
\newblock Large feedback arc sets, high minimum degree subgraphs, and long
  cycles in {E}ulerian digraphs.
\newblock {\em Combinatorics, Probability and Computing}, 22(6):859--873, 2013.

\bibitem[IK99]{IshaiK99}
Yuval Ishai and Eyal Kushilevitz.
\newblock Improved upper bounds on information-theoretic private information
  retrieval (extended abstract).
\newblock In {\em Proceedings of the Thirty-First Annual {ACM} Symposium on
  Theory of Computing, May 1-4, 1999, Atlanta, Georgia, {USA}}, pages 79--88.
  {ACM}, 1999.

\bibitem[Jan23]{Jan23}
Oliver Janzer.
\newblock Rainbow {T}ur{\'a}n number of even cycles, repeated patterns and
  blow-ups of cycles.
\newblock {\em Israel Journal of Mathematics}, 253(2):813--840, 2023.

\bibitem[JM25]{JanzerM25}
Oliver Janzer and Peter Manohar.
\newblock A $k^{\frac{q}{q-2}}$ lower bound for odd query locally decodable
  codes from bipartite kikuchi graphs.
\newblock In {\em 66th {IEEE} Annual Symposium on Foundations of Computer
  Science, {FOCS} 2025, Sydney, Australia, December 14-17, 2025}, pages 75--85.
  {IEEE}, 2025.

\bibitem[JS24]{JS24}
Oliver Janzer and Benny Sudakov.
\newblock On the {T}ur{\'a}n number of the hypercube.
\newblock In {\em Forum of Mathematics, Sigma}, volume~12, page e38. Cambridge
  University Press, 2024.

\bibitem[KLLT24]{KLLT24}
Jaehoon Kim, Joonkyung Lee, Hong Liu, and Tuan Tran.
\newblock Rainbow cycles in properly edge-colored graphs.
\newblock {\em Combinatorica}, 44(4):909--919, 2024.

\bibitem[KM24a]{KothariM23}
Pravesh~K. Kothari and Peter Manohar.
\newblock An exponential lower bound for linear 3-query locally correctable
  codes.
\newblock In {\em Proceedings of the 56th Annual {ACM} Symposium on Theory of
  Computing, {STOC} 2024, Vancouver, BC, Canada, June 24-28, 2024}, pages
  776--787. {ACM}, 2024.

\bibitem[KM24b]{KothariM24}
Pravesh~K. Kothari and Peter Manohar.
\newblock Exponential lower bounds for smooth 3-lccs and sharp bounds for
  designs.
\newblock In {\em 65th {IEEE} Annual Symposium on Foundations of Computer
  Science, {FOCS} 2024, Chicago, IL, USA, October 27-30, 2024}. {IEEE}, 2024.

\bibitem[KMRS16]{KoppartyMRS16}
Swastik Kopparty, Or~Meir, Noga Ron{-}Zewi, and Shubhangi Saraf.
\newblock High-rate locally-correctable and locally-testable codes with
  sub-polynomial query complexity.
\newblock In {\em Proceedings of the 48th Annual {ACM} {SIGACT} Symposium on
  Theory of Computing, {STOC} 2016, Cambridge, MA, USA, June 18-21, 2016},
  pages 202--215. {ACM}, 2016.

\bibitem[KMSV07]{KMSV04}
Peter Keevash, Dhruv Mubayi, Benny Sudakov, and Jacques Verstra{\"e}te.
\newblock Rainbow {T}ur{\'a}n problems.
\newblock {\em Combinatorics, Probability and Computing}, 16(1):109--126, 2007.

\bibitem[KSY11]{KoppartySY11}
Swastik Kopparty, Shubhangi Saraf, and Sergey Yekhanin.
\newblock High-rate codes with sublinear-time decoding.
\newblock In {\em Proceedings of the 43rd {ACM} Symposium on Theory of
  Computing, {STOC} 2011, San Jose, CA, USA, 6-8 June 2011}, pages 167--176.
  {ACM}, 2011.

\bibitem[KW04]{KerenidisW04}
Iordanis Kerenidis and {Ronald de} Wolf.
\newblock Exponential lower bound for 2-query locally decodable codes via a
  quantum argument.
\newblock {\em Journal of Computer and System Sciences}, 69(3):395--420, 2004.

\bibitem[LFKN90]{LundFKN90}
Carsten Lund, Lance Fortnow, Howard~J. Karloff, and Noam Nisan.
\newblock Algebraic methods for interactive proof systems.
\newblock In {\em 31st Annual Symposium on Foundations of Computer Science, St.
  Louis, Missouri, USA, October 22-24, 1990, Volume {I}}, pages 2--10. {IEEE}
  Computer Society, 1990.

\bibitem[RS96]{RubinfeldS96}
Ronitt Rubinfeld and Madhu Sudan.
\newblock Robust characterizations of polynomials with applications to program
  testing.
\newblock {\em SIAM Journal on Computing}, 25(2):252--271, 1996.

\bibitem[Sha90]{Shamir90}
Adi Shamir.
\newblock Ip=pspace.
\newblock In {\em 31st Annual Symposium on Foundations of Computer Science, St.
  Louis, Missouri, USA, October 22-24, 1990, Volume {I}}, pages 11--15. {IEEE}
  Computer Society, 1990.

\bibitem[Tom24]{Tom24}
Istv{\'a}n Tomon.
\newblock Robust (rainbow) subdivisions and simplicial cycles.
\newblock {\em Advances in Combinatorics}, 2024:1, 37 pp, 2024.

\bibitem[Tre04]{Trevisan04}
Luca Trevisan.
\newblock Some applications of coding theory in computational complexity.
\newblock {\em arXiv preprint cs/0409044}, 2004.

\bibitem[Yan24]{Yankovitz24}
Tal Yankovitz.
\newblock A stronger bound for linear 3-lcc.
\newblock In {\em 65th {IEEE} Annual Symposium on Foundations of Computer
  Science, {FOCS} 2024, Chicago, IL, USA, October 27-30, 2024}. {IEEE}, 2024.

\bibitem[Yek12]{Yekhanin12}
Sergey Yekhanin.
\newblock Locally decodable codes.
\newblock {\em Foundations and Trends in Theoretical Computer Science},
  6(3):139--255, 2012.

\end{thebibliography}

\appendix

\end{document}